\documentclass[a4paper, 11pt]{article} 
\def\eps{png} % *ja* fuer eps-Bilder oder *etwas anders* fuer png-Bilder 
\pdfoutput=1
\usepackage[T1]{fontenc}
\usepackage{ifthen}
\usepackage{rotating}
\usepackage[all]{xy}
\usepackage{amssymb}
\usepackage{amsfonts}
\usepackage{amsthm}
\usepackage{amsmath}
\usepackage{graphics}
\usepackage{hyperref}
\hypersetup{colorlinks=true,citecolor=black}
\usepackage{srcltx}
\graphicspath{{Graphics/}}
\newtheorem{Thm}{Theorem}{\bf}{\it}
\newtheorem{Pro}[Thm]{Proposition}{\bf}{\it}
\newtheorem{Cor}[Thm]{Corollary}{\bf}{\it}
\newtheorem{Lem}[Thm]{Lemma}{\bf}{\it}
\theoremstyle{definition}
\newtheorem{Rem}[Thm]{Remark}{\it}{\rm}
\newtheorem{Exa}[Thm]{Example}{\it}{\rm}
\newtheorem{Def}[Thm]{Definition}{\bf}{\rm}
\newcommand{\bC}{\mathbb{C}}
\newcommand{\bE}{\mathbb{E}}
\newcommand{\bG}{\mathbb{G}}

\newcommand{\bM}{\mathbb{M}}
\newcommand{\bP}{\mathbb{P}}
\newcommand{\bR}{\mathbb{R}}
\newcommand{\cA}{\mathcal{A}}
\newcommand{\cB}{\mathcal{B}}
\newcommand{\cE}{\mathcal{E}}

\newcommand{\cG}{\mathcal{G}}

\newcommand{\cS}{\mathcal{S}}
\newcommand{\tr}{{\rm tr}}
\newcommand{\sa}{\cA_{{\rm sa}}}
\newcommand{\st}{\cS}

\newcommand{\lin}{{\rm lin}}

\newcommand{\ri}{{\rm ri}}

\newcommand{\id}{{\rm 1\mskip-4mu l}}

\newcommand{\winkel}{\angle\begin{picture}(0,0)%
\put(-0.25,-0.06){\oval(0.35,0.35)[tr]}\end{picture}}
\newcommand{\bsm}{\left(\begin{smallmatrix}} % begin small matrix
\newcommand{\esm}{\end{smallmatrix}\right)}  % end   small matrix
\usepackage{enumerate}
\begin{document}
\begin{center} 
\textbf{\Large{Entropy Distance: New Quantum Phenomena}}\\[5mm]
Stephan Weis\footnote{\texttt{sweis@mis.mpg.de}}\\
Max Planck Institute for Mathematics in the Sciences\\
Inselstr.\ 22, D-04103 Leipzig,
Germany\\[3mm]
and\\[3mm]
Andreas Knauf\footnote{\texttt{knauf@mi.uni-erlangen.de}}\\
Department of Mathematics,
Friedrich-Alexander-University Erlangen-Nuremberg,
Cauerstr.\ 11, D-91058 Erlangen, 
Germany\\[5mm]
September 5, 2012
\end{center}
\begin{abstract}
We study a curve of Gibbsian families of complex
$3\times 3$-matrices and point out new features, absent in 
commutative finite-dimensional algebras: a discontinuous 
maximum-entropy inference, a discontinuous entropy distance
and non-exposed faces of the mean value 
set. We analyze these problems from various aspects including
convex geometry, topology and information geometry. This research
is motivated by a theory of infomax principles, where we 
contribute by computing first order optimality conditions
of the entropy distance.
\end{abstract}
%
%
%
%-----------------------------------------------------------
\section{Introduction}
%-----------------------------------------------------------
%
%
%
%
\par
The aim of the introduction is a discussion of the maximum-entropy 
inference under linear constraints, in two aspects: The problem of its 
discontinuity and its connection to infomax principles, asking for 
maximization of the entropy distance from an exponential family. 
Section~\ref{sec:summary} gives an overview of the article.%}
\subsection{Maximum-entropy inference and infomax principles}
\label{max_inf}
\par
The \emph{maximum-entropy principle}, while dating back to Boltzmann,
became the information theoretic justification of the thermodynamic 
formalism, see \cite{Jaynes1}. We have discovered in 
three-level quantum systems a problem that can arise for 
non-commutative observables: The real analytic maximum-entropy 
inference under linear constraints has no continuous extension. 
An example is given in Remark~\ref{rem:max_entropy_family}; this 
phenomenon does not appear in commutative algebras of finite 
dimensions. 
\par
The roughness of a discontinuity in the maximum-entropy 
inference shows that we are currently at the very beginning of a 
quantitative understanding of its performance. A deeper
analysis seems necessary to tackle applications based on
asymptotic statistical variance
or on asymptotic error rates. 
Other branches of quantum inference, e.g.\ state tomography
\cite{Wootters,Petz_Ruppert} or hypothesis 
testing \cite{Audenaert,Nussbaum}, are further developed and 
asymptotic error rates are used to identify 
optimal tests.
\par
What do we mean by a discontinuous maximum-entropy inference? 
We use a fixed set of \emph{observables} $a_1,\ldots,a_k$, 
i.e.\ self-adjoint matrices in the algebra $\mathcal A={\rm Mat}(N,\bC)$, 
and denote by ${\cal A}_{\rm sa}$ the real vector space of self-adjoint 
matrices. We assume a quantum system is described by a 
\emph{density matrix} $\rho$, also called \emph{state}, i.e.\ 
$\rho\in{\rm Mat}(N,\bC)$ ($N$-level system), $\rho\succeq 0$ 
(positive semi-definite) and ${\rm tr}(\rho)=1$ (normalized). 
We denote by ${\cal S}({\cal A})$ the set of density matrices, 
called \emph{state space}. 
We assume a generic quantum systems where the
density matrix $\rho$ is invertible.
\par
The von Neumann measurements (see \cite{Petz}) of 
$a_r=\sum_{\lambda\in{\rm spec}(a_r)}\lambda P_{r,\lambda}$ yield 
eigenvalue $\lambda$ with probability $\tr(\rho P_{r,\lambda})$.
\begin{itemize}
\item[$\bullet$]
If $n$ copies of $\rho$ are available for measurement (in form of the 
$n$-fold tensor product 
$\rho\otimes\cdots\otimes\rho\in {\cal A}_{\rm sa}^{\otimes n}$), 
then $n$ measurements of $a_r$ give us eigenvalues 
$\lambda_1,\ldots,\lambda_n\in{\rm spec}(a_r)$ such that the mean
\begin{equation}
\label{eq:empirical_mean}\textstyle
\overline{a_r}(n)
\;:=\;
\tfrac 1{n}(\lambda_1+\ldots+\lambda_n)
\end{equation}
converges to the true mean ${\rm tr}(\rho a_r)$ by the strong law 
of large numbers. 
\item[$\bullet$]
If $nk$ copies of $\rho$ are available,
the measured values $m_1,\ldots, m_k$ of the 
$k$ random variables $\overline{a_1}(n),\ldots,\overline{a_k}(n)$ 
define an affine subspace 
\begin{equation}
\label{eq:empiric}\textstyle
\big\{\sigma
\in{\cal A}_{\rm sa}\mid 
\big({\rm tr}(\sigma a_1),\ldots,{\rm tr}(\sigma a_k)\big)
\;=\;
\big(m_1,\ldots,m_k\big)\,\big\}.
\end{equation}
\par
We assume that this subspace intersects the state space 
${\cal S}({\cal A})$,
since by large deviation theory  (e.g., Chap. I.3 of \cite{Ellis})
the probability of a distance 
larger than a given $\varepsilon>0$ from 
{$\rho$}
decays exponentially in $n$.
\end{itemize}
The {\it maximum-entropy inference} associates to the measured values
$\big(m_1,\ldots,m_k\big)$ the unique density matrix $\widehat{\rho}(n)$ 
in the set of states $\sigma$ satisfying (\ref{eq:empiric}) which 
maximizes the \emph{von Neumann entropy} 
\begin{equation}
\label{eq:maxent_inference}
S(\sigma)\;:=\;-{\rm tr}(\sigma\log(\sigma))\,.
\end{equation}
\par
The maximum-entropy inference is well-defined since the von Neumann
entropy is a strictly concave function \cite{Wehrl}. The inference
is a real analytic mapping on the domain of all mean value tuples 
$\big({\rm tr}(\sigma a_1),\ldots,{\rm tr}(\sigma a_k)\big)$
for invertible density matrices $\sigma$, see e.g.\ \cite{Wichmann}. 
The image, called \emph{Gibbsian family} (of density matrices),  
consists of all matrices of the form
\[
\exp(a_0+\lambda_1a_1+\cdots+\lambda_ka_k)
/{\rm tr}(\exp(a_0+\lambda_1a_1+\cdots+\lambda_ka_k))
\]
for real $\lambda_1,\ldots,\lambda_k$ and $a_0=0$. In general,
if $a_{r}\in{\cal A}_{\rm sa}$, this manifold of density
matrices is called \emph{exponential family}.
\par
In Remark~\ref{rem:max_entropy_family} we discuss a Gibbsian family 
where the real analytic maximum-entropy inference 
defined on the interior of the mean value set has no continuous
extension to the full mean value set.
While the variance of the random variables $\overline{a_r}(n)$, 
$r=1,\ldots,k$ in (\ref{eq:empirical_mean}) and of the tuple
$(\overline{a_1}(n),\ldots,\overline{a_k}(n))$ is ${\cal O}(1/n)$, 
the statement is not obvious for the maximum-entropy inference
$\widehat{\rho}(n)$. Indeed, the lack of continuous extension
shows that the constant in the variance estimate ${\cal O}(1/n)$ of 
$\widehat{\rho}(n)$ can be arbitrarily large.
For the non-generic choice of a singular density matrix 
$\rho$ the limit $\lim_{n\to\infty}\widehat{\rho}(n)$ needs not even
be a state of maximum von Neumann entropy. 
Convergence rates of the maximum-entropy inference
were considered in the context of model selection \cite{Rau}.
\par
Maximum-entropy inference is closely connected to the entropy distance
from an exponential family.
The \emph{relative entropy} between states $\rho,\sigma\in\st(\cA)$ is 
$S(\rho,\sigma):=+\infty$ unless the image of $\sigma$ contains 
that of $\rho$ and then (using the natural logarithm)
\begin{equation}
\label{eq:rel_ent}
S(\rho,\sigma)\;:=\;\tr\,\rho\big(\ln(\rho)-\ln(\sigma)\big)\,.
\end{equation}
The distance-like properties of $S(\rho,\sigma)\geq 0$ and of 
$S(\rho,\sigma)=0\iff\rho=\sigma$ hold \cite{Wehrl}.
However, the relative entropy is not a metric.
For $\cE\subset \st(\cA)$ 
\begin{equation}
\label{eq:entropy_distance_def}\textstyle
{\rm d}_\cE\;:\quad\st(\cA)\,\rightarrow\,\bR,\quad
\rho\,\mapsto\,\inf_{\sigma\in\cE}S(\rho,\sigma)\,
\end{equation}
is called \emph{entropy distance} 
of $\rho$ from $\cE$. If $\cE$ contains invertible density matrices, 
then ${\rm d}_\cE$ is bounded on $\st(\cA)$.
\par
Under arbitrary constraints, maximizing the von Neumann entropy  is 
the same as minimizing the relative entropy distance 
${\rm d}_{\{\id/{\rm tr}(\id)\}}$
from the tracial state. 
In Section~\ref{subsec:why_mean} we recall 
that for linear constraints the latter is equivalent to the unconstrained 
minimization 
of the relative entropy in its second argument from the corresponding 
Gibbsian family.
\par
Infomax principles support the hypothesis that natural systems tend 
to maximize structured correlations. This, in the work \cite{Ay}, is 
formalized as deviation from an exponential family $\cE$, and is
quantified by the entropy distance (\ref{eq:entropy_distance_def}). 
An instructive example is the mutual information used in information 
theory: 
\begin{Exa}[Product States]
The \emph{mutual information} of a bipartite state $\rho_{AB}$ is given by 
$S(\rho_{AB},\rho_A\otimes\rho_B)\geq 0$ for the relative entropy $S$
and for reduced states $\rho_A$ resp.\
$\rho_B$ on subsystem $A$ resp.\ $B$. It is zero only when
$\rho_{AB}=\rho_A\otimes\rho_B$. The relative entropy measures
the distance of an arbitrary bipartite state from the 
Gibbsian family of all product states. 
\end{Exa}
The mutual information of a quantum system measures the total 
correlation of a bipartite quantum system.
For the entanglement in the system there exist other measures, e.g.\
the entropy distance from the set of separable states, known as 
{\it relative entropy of entanglement}, see e.g.\ \cite{Vedral}. 
Correlation measures having the form of the entropy distance from a 
Gibbsian family are used in statistical physics, image processing or 
in the theory of neural networks to just name a few, see e.g.\ 
\cite{Matsuda,Erb,amari_hierarchy,Ay_Jost}. 
\par
Maximizers of the entropy distance from an exponential families 
(of probability distributions)
were studied e.g.\ in \cite{Ay,Knauf,Matus07,Rauh,Matus_Rauh}. 
In Section~\ref{sec:maxi} we contribute to a non-commutative
analogon by computing first order optimality conditions.
%
%
%
%
%
%
%%%%%%%%%%%%%%%%%%%%%%%%%%%%%%%%%%%%%%%%%%%%%%%%%%%%%%%%%%%%%
%%%%%%%%%%%%%%%%%%%%%%%%%%%%%%%%%%%%%%%%%%%%%%%%%%%%%%%%%%%%%
%
\subsection{Summary of our results}
\label{sec:summary}
\par
Most of the rest of the paper will focus on observables in the
algebra of Example~\ref{exa:algebra}.
We study a curve of planes in a Grassmannian 
manifold of linear spaces 
that defines a curve of two-dimensional Gibbsian families of 
$3\times 3$-density matrices. Unlike Gibbsian 
families in finite probability spaces, one of the families has 
a discontinuous entropy distance and its real analytic 
maximum-entropy inference does not extend continuously. We 
discuss several candidates of closures to extend Gibbsian 
families and we propose a convex geometric criterion to
characterize discontinuities: Where non-exposed faces are 
born in a Grassmannian manifold of linear subspaces, families 
have a discontinuous inference.
This conjecture is supported by the example
of the Staffelberg family in Section~\ref{sec:staffelberg}.
\par
To compare classical and quantum physics, we consider 
{\it *-subalgebras} ${\cal A}$ of ${\rm Mat}(N,\mathbb C)$. To allow 
low-dimensional examples we consider them {\it real}, i.e.\ 
${\cal A}$ is a subring of ${\rm Mat}(N,\mathbb C)$, and an $\bR$-module 
closed under conjugation $a\mapsto a^*$. However, it is not necessarily 
closed under complex scalar multiplication. 
The {\it state space} of ${\cal A}$ is the set 
${\cal S}={\cal S}({\cal A})=
\{\rho\in\mathcal A\mid\rho\succeq 0,{\rm tr}(\rho)=1\}$
of density matrices. 
We denote $\id$ / $0$ resp.\ $\id_N$ / $0_N$ the identity / zero in 
${\cal A}$ resp.\ ${\rm Mat}(N,\mathbb C)$. We allow for $\id\neq\id_N$ 
which we need to study the {\it swallow family} in Section~\ref{sec:swallow}
and to prove an optimality condition in Section~\ref{sec:maxi}, see
also Remark~\ref{rem:supp_extend}.
The real vector space of self-adjoint matrices ${\cal A}_{\rm sa}$ is a 
Euclidean vector space for the {\it Hilbert-Schmidt} scalar 
product $\langle a,b\rangle={\rm tr}(ab)$, $a,b\in{\cal A}_{\rm sa}$.
\begin{Rem}
There are other natural definitions of the state space of a real 
*-subalgebra $\mathcal A$ of ${\rm Mat}(N,\mathbb C)$, e.g.\ 
\begin{enumerate}
\item
the density matrices in $\mathcal A$ (like above),
\item
the states on ${\rm Mat}(N,\mathbb C)$ restricted to
$\mathcal A_{\rm sa}$,
\item
the positive linear functionals on $\mathcal A_{\rm sa}$ that take 
the value $1$ at the identity.
\end{enumerate}
These definitions are mutually equivalent, assuming $\id_N\in\mathcal A$.
The inclusions of 1.\ into 2.\ into 3.\ are trivial.
The inclusion of 3.\ into 2.\ follows from the Riesz extension theorem 
and the inclusion of 2.\ into 1.\ follows from the fact that orthogonal 
projection from ${\rm Mat}(N,\mathbb C)_{\rm sa}$ onto 
$\mathcal A_{\rm sa}$ takes density matrices to density matrices.
\end{Rem}
\par
The following real *-subalgebra of the C*-algebra 
${\rm Mat}(2,{\mathbb C})\oplus\bC$ is sufficiently rich for our 
purposes and it includes the curve of Gibbsian families. The state 
space of ${\rm Mat}(2,{\mathbb C})\oplus\bC$ has already been analyzed 
in \cite{Kuperberg} as the simplest example of a 'hybrid' memory (and 
called {\em hybrid trit}) but the main subject of that article is not 
relevant to our discussions.
\begin{Exa}
\label{exa:algebra}
We consider the real *-subalgebra 
$\mathcal B\subset{\rm Mat}(2,\mathbb C)$ spanned by
$\id_2$, $\sigma_1$, $\sigma_2$ and  ${\rm i}\sigma_3$
for the {\it Pauli $\sigma$-matrices}
$\sigma_1:=\left(\begin{smallmatrix}0&1\\1&0\end{smallmatrix}\right)$,
$\sigma_2:=\left(\begin{smallmatrix}0&-{\rm i}\\{\rm i}&0
\end{smallmatrix}\right)$,
$\sigma_3:=\left(\begin{smallmatrix}1&0\\0&-1\end{smallmatrix}\right)$.
This algebra is isomorphic to ${\rm Mat}(2,\mathbb R)$ by exchanging
$\sigma_2$ and $\sigma_3$. 

A real *-subalgebra
$\mathcal A\subset{\rm Mat}(3,\mathbb C)$ is defined by
block diagonal matrices 
$\left(\begin{smallmatrix}*&*&0\\{}*&*&0\\0&0&*\end{smallmatrix}\right)$
with elements of $\cB$ in the upper left corner and real numbers in the 
lower right corner. The state space of $\mathcal B$ is 
$\st(\cB)={\rm conv}\{\tfrac 1{2}
(\id_2+\sin(\alpha)\sigma_1+\cos(\alpha)\sigma_2)
\mid\alpha\in\bR\}$ where ${\rm conv}$ denotes convex hull. This
disk is a section of the state space of ${\rm Mat}(2,\mathbb C)$, 
known as {\it Bloch ball}. The state space of $\cA$ is a three-dimensional 
cone based on $\mathcal S(\mathcal B)\oplus 0$ and with apex $0_2\oplus 1$,
\[
\st(\cA)={\rm conv}(0_2\oplus 1, \rho(\alpha); \alpha\in\bR)
\]
for 
\[
\rho(\alpha)
\;:=\;
\tfrac 1{2}(\id_2+\sin(\alpha)\sigma_1+\cos(\alpha)\sigma_2)\oplus 0\,.
\]
It is the solid of revolution of an equilateral triangle.
\end{Exa}
\par
It is well known that state spaces of commutative and non-commutative
algebras have quite different geometries. Whereas in the commutative
case we have a simplex (and thus every state is uniquely decomposed into 
pure states), in the non-commutative case such a decomposition is highly 
non-unique (think of the Bloch ball).
\begin{figure}
\begin{center}
\begin{picture}(8,3)
\ifthenelse{\equal{\eps}{ja}}{%
\put(0,0){\includegraphics[height=3cm, bb=200 160 560 520, clip=]%
{Fig11.eps}}
\put(5,0){\includegraphics[height=3cm, bb=140 160 520 540, clip=]%
{Fig12.eps}}}{%
\put(0,0){\includegraphics[height=3cm, bb=200 160 560 520, clip=]%
{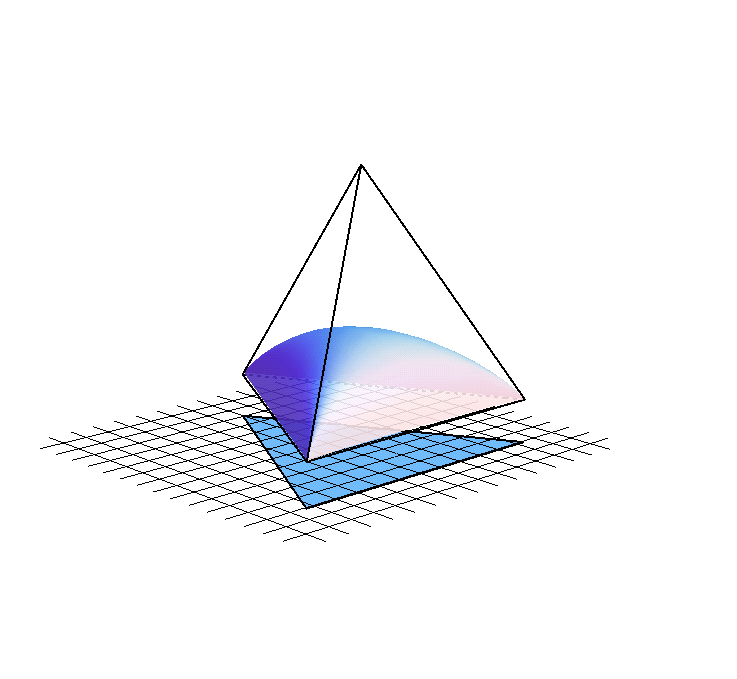}}
\put(5,0){\includegraphics[height=3cm, bb=140 160 520 540, clip=]%
{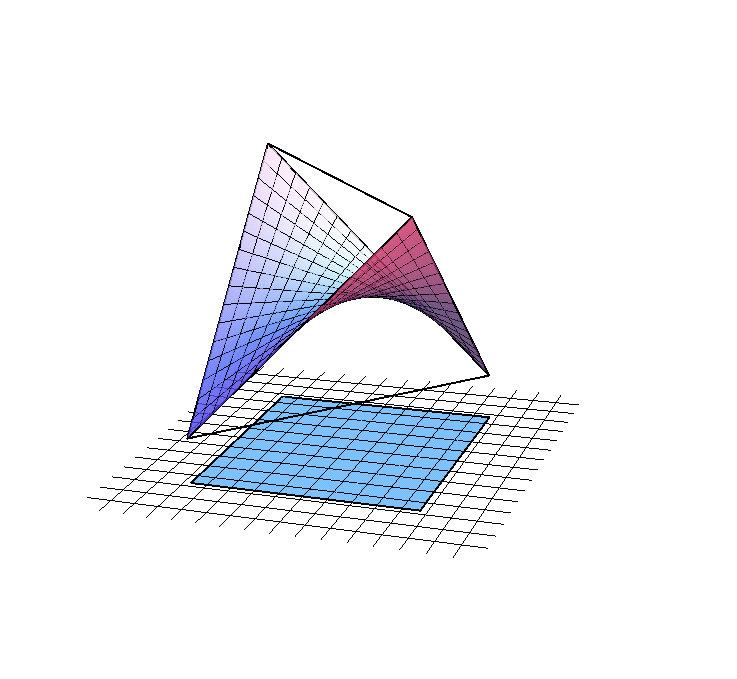}}}
%
%\put(0,0){\line(1,0){5}}
%\put(0,5){\line(1,0){5}}
%\put(0,0){\line(0,1){5}}
%\put(5,0){\line(0,1){5}}
%
%\put(7,0){\line(1,0){5}}
%\put(7,5){\line(1,0){5}}
%\put(7,0){\line(0,1){5}}
%\put(12,0){\line(0,1){5}}
\end{picture}
\caption{\label{fig:families}Mean value sets for two probabilistic 
exponential families. Left: triangle; right: square.}
\label{fig:Meanvaluesets}
\end{center}
\end{figure}
\par
Still, from the point of view of convex geometry there is one 
common property of all these state spaces: all of their faces are
exposed, that is, they can be described as the intersection of 
state space with a half space.
{\em Non-exposed faces} are found, 
e.g., on the circumference of a stadium, at the four points where 
a half-circle meets a segment. See Section~\ref{sec:cone_family} 
for precise definitions.
In the probabilistic setting of
$\cA\cong\bC^N$, embedded as diagonal matrices, measurement of 
observables $f_1,\ldots,f_n$ leads to an orthogonal projection 
\[
\st(\cA)\,\longrightarrow\,\bR^n,
\quad 
p\,\longmapsto\,\big(\bE_p(f_1),\ldots,\bE_p(f_n)\big)
\]
of state space, based on expectation $\bE_p$. The image, called 
\emph{mean value set} or \emph{convex support} 
\cite{Barndorff} is no longer a 
simplex but still a polygon. So faces of a mean value set 
are exposed, too. The same applies to all exponential families and 
their natural projections, see Figure \ref{fig:Meanvaluesets}. 
\begin{figure}
\begin{center}
\begin{picture}(12,5)
\ifthenelse{\equal{\eps}{ja}}{%
\put(0,0){\includegraphics[height=5cm, bb=0 0 500 500, clip=]%
{qm_projections.eps}}
\put(7,0){\includegraphics[height=5cm, bb=0 0 500 500, clip=]%
{qm_sections.eps}}}{%
\put(0,0){\includegraphics[height=5cm, bb=0 0 500 500, clip=]%
{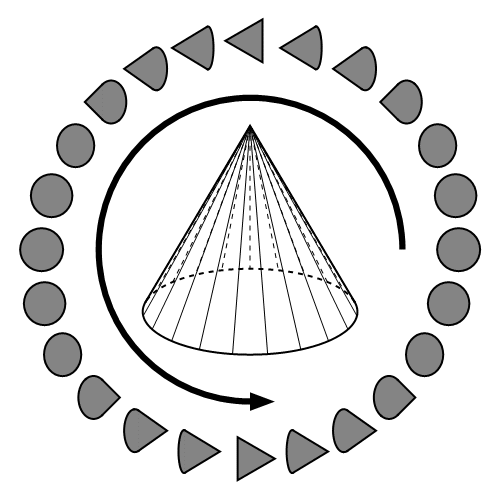}}
\put(7,0){\includegraphics[height=5cm, bb=0 0 500 500, clip=]%
{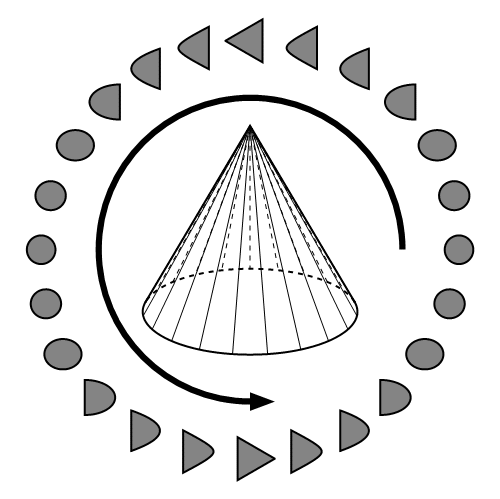}}}
%
%\put(0,0){\line(1,0){5}}
%\put(0,5){\line(1,0){5}}
%\put(0,0){\line(0,1){5}}
%\put(5,0){\line(0,1){5}}
%
%\put(7,0){\line(1,0){5}}
%\put(7,5){\line(1,0){5}}
%\put(7,0){\line(0,1){5}}
%\put(12,0){\line(0,1){5}}
\end{picture}
\caption{\label{fig:reflections}
The 3D cone is the state space of a non-commutative algebra.
Left: Mean value sets (projections of the cone); right: 
sections of the cone. Projections respectively sections are w.r.t.\ 
planes including the tracial state $\id/{\rm tr}(\id)$, which is the
centroid of the cone.
}
\end{center}
\end{figure}
\par
We exhibit here two main differences between exponential families 
in commutative and non-commutative algebras, at least in 
the curve of our example. 
\begin{itemize}
\item[$\bullet$]
First, we show in Section~\ref{sec:cone_family} that it is 
typical for a non-commutative algebra that mean value sets
have non-exposed faces.
\item[$\bullet$]
Second, we show in Section~\ref{sec:staffelberg} that
the entropy distance from an exponential family can be 
discontinuous in exceptional cases. 
\end{itemize}
In Figure~\ref{fig:reflections} (left) we have sketched two-dimensional 
mean value sets of the state space $\st(\cA)$ from 
Example~\ref{exa:algebra}. A mean value set has non-exposed faces if it 
is the convex hull of a non-degenerate ellipse and of an exterior point. 
Mean value sets with non-exposed faces are bounded in the Grassmannian 
manifold by elliptical shapes that correspond to exponential families 
with discontinuous entropy distance (angles 
$\tfrac{1}{6}\pi,\tfrac{5}{6}\pi,\tfrac{7}{6}\pi$ and $\tfrac{11}{6}\pi$).
It seems this boundary in the Grassmannian manifold is pivotal for 
discontinuity. 
\par
Towards a classification of non-exposed faces of mean value sets
one can study singularities of the dual convex set (see \cite{Weis_dual},
including references on the progress of that question). These dual
convex sets are sections of the state space (cf.\ \cite{Weis_supp} and 
Figure~\ref{fig:reflections}, right) and they are bounded by 
determinantal varieties which are a subject of study in convex algebraic 
geometry, see e.g.\ \cite{Netzer}.
\par
Already in 1963 ensembles of maximum chaos in $\cA={\rm Mat}(N,\bC)$
were studied in \cite{Wichmann}. However, non-exposed faces at a
mean value set have attracted little attention in the literature.
In particular Theorem I (e) in \cite{Wichmann}, concerning
extreme points is wrong, it fails in all cases where non-exposed
extreme points appear. An example is given in 
Remark~\ref{rem:swallow} a).
We are convinced that non-exposed faces are important in the analysis 
of maximum-entropy inference and entropy distance. 
As we have seen in the beginning of this section the 
convex geometric notion of non-exposed face indicates discontinuity of 
the inference.
\par
We will show later in this paper that the maximum-entropy 
inference does not extend continuously. So the
question arises how a Gibbsian family ${\cal G}$ must be
extended to a locus of maximum-entropy density matrices under linear
constraints.
It is clear that the topological norm closure is too 
large. In the examples presented  
in Section~\ref{sec:closures}
we will prove that the 
\emph{reverse information closure} or \emph{rI-closure}
\begin{equation}
\label{eg:rI_closure}
{\rm cl}^{\rm rI}({\cal G})
\;:=\;\{\rho\in\st(\cA)\mid \inf_{\sigma\in{\cal G}}S(\rho,\sigma)=0\}
\end{equation}
gives the right answer. Its name is motivated from probability
theory \cite{Csiszar03} and it consists of states that approximate 
${\cal G}$ in relative entropy $S$. Since the algebra ${\cal A}$ is
a nice substructure of
${\rm Mat}(N,\mathbb C)$, we have also a theory
of information geometry \cite{Amari} at our disposition, which gives 
us two canonical choices of geodesics on the manifold ${\cal G}$.
These $(+1)$-geodesics and $(-1)$-geodesics will be defined in the 
next section. They give rise to the 
\emph{$(+1)$-closure}
\begin{equation}
\label{eg:1_closure}
{\rm cl}^{(+1)}({\cal G})\;:=\;{\cal G}
\;\cup\;\{\;\text{limit points of 
$(+1)$-geodesics in }{\cal G}\;\}
\end{equation}
and the \emph{$(-1)$-closure}
\begin{equation}
\label{eg:-1_closure}
{\rm cl}^{(-1)}({\cal G})\;:=\;{\cal G}\;\cup\;
\{\;\text{limit points of $(-1)$-geodesics in }{\cal G}\;\}\,.
\end{equation}
The inclusions of ${\rm cl}^{(+1)}({\cal G})$ and 
${\rm cl}^{(-1)}({\cal G})$ into the norm closure $\overline{\cal G}$
are obvious. We show
\begin{equation}
\label{eq:closure_inclusion}
{\rm cl}^{(+1)}({\cal G})
\;\subset\;
{\rm cl}^{\rm rI}({\cal G})
\;\subset\;
\overline{\cal G}\,,
\end{equation}
where the second inclusion follows from the Pinsker-Csisz\'ar
inequality. We prove that the $(+1)$-closure is
smaller than the locus of maximum-entropy density matrices, 
and that the rI-closure and the $(-1)$-closure are possible
candidates for the correct extension of $\cG$.
%
%
%
%
%
%
%
%%%%%%%%%%%%%%%%%%%%%%%%%%%%%%%%%%%%%%%%%%%%%%%%%%%%%%%%%%%%%
%%%%%%%%%%%%%%%%%%%%%%%%%%%%%%%%%%%%%%%%%%%%%%%%%%%%%%%%%%%%%
%%%%%%%%%%%%%%%%%%%%%%%%%%%%%%%%%%%%%%%%%%%%%%%%%%%%%%%%%%%%%
%
%
%
%
\section{Two affine charts and some remarks}
\label{subsec:why_mean}
\par
We introduce two sorts of canonical geodesics on a Gibbsian 
family and we provide a geometric discussion of how the 
maximum-entropy inference relates to the entropy distance.
We remark on the information geometric context of the geodesics,
on quantum channels and on advantages of real *-subalgebras as 
opposed to C*-subalgebras.
\par
In this section $\cA$ denotes an arbitrary real *-subalgebra
of ${\rm Mat}(N,\mathbb C)$.
The set of invertible states equals the relative interior
of the state space
\[\textstyle
\ri\,\st(\cA)\;=\;
\{\rho\in\st(\cA)\mid\rho^{-1}\;\textrm{exists in}\;\cA\}\,,
\]
i.e.\ the interior of $\st(\cA)$ in its affine span
${\cal A}_1:=\{a\in{\cal A}_{\rm sa}\mid{\rm tr}(a)=1\}$,
see e.g.\ Proposition 2.9 in \cite{Weis_supp}.
The trace-normalized exponential is the real analytic mapping
\[\textstyle
\exp_1\;:\quad
\sa\,\longrightarrow\,\ri\,\st(\cA),\quad
a\,\longmapsto\,\frac{e^a}{\tr(e^a)}
\]
defined by functional calculus of self-adjoint matrices in $\cA$.
This is a diffeomorphism when restricted to traceless 
matrices. The real analytic inverse 
$\ln_0: \ri(\cS(\cA)) \to \cA_0$ defined by
\[\textstyle
\ln_0:\rho\mapsto\ln(\rho)-\id\,\tr(\ln(\rho))/\tr(\id)
\]
is the \emph{canonical chart} of $\ri\,\cS(\cA)$. 
\par
The image of a non-empty
affine subspace of $\sa$ under $\exp_1$ is an \emph{exponential family}
in $\cA$. For an exponential family $\cE$ we call $\ln_0|_\cE$ the 
\emph{canonical chart} of $\cE$. The affine space 
$\Theta:=\ln_0(\cE)$ is the \emph{canonical parameter space},
its translation vector space $V:=\{x-y\mid x,y\in \Theta\}$ is the 
\emph{canonical tangent space} and the restriction of $\exp_1$ to 
$\Theta$ is the \emph{canonical parametrization} of $\cE$.
\par
An exponential family is a Gibbsian family if $\Theta=V$, and for that
case a different chart was introduced in Theorem 2 (b) in \cite{Wichmann}:
If $\pi_V:{\cal A}_{\rm sa}\to V$ denotes orthogonal projection onto 
$V$, we define the {\it mean value set}
\begin{equation}
\label{eq:mean_value_set}
\bM(V)\;=\;\bM_\cA(V)\;:=\;\pi_V\big(\st(\cA)\big)\,.
\end{equation}
The mean value set is affinely isomorphic to
$\{(\langle \rho,v_1\rangle,\ldots,\langle \rho,v_1\rangle\mid
\rho\in\st(\cA)\}$, if  $v_1,\ldots,v_k$ is a basis of $V$, see 
e.g.\ Remark~1.1 in \cite{Weis_supp}. The latter set was used in 
\cite{Wichmann}. It is not reasonable to choose a basis of $V$ 
in our analysis, because vector spaces $pVp$ for
projections $p=p^2=p^*\in\cA$ will be used, see 
Remark~\ref{rem:supp_extend}, and multiplication with $p$ can
destroy linear independence.

The map $\pi_V\circ\exp_1|_V:V\to\ri\big({\bM}(V)\big)$ is
a real analytic diffeomorphism, its image is an open subset
of $V$. The \emph{mean value chart} for the Gibbsian family
$\cE$ is the bijection
\begin{equation}
\label{eq:mean_chart}
\pi_V|_{\cE}\;:\quad
\cE\,\longrightarrow\,\ri\big({\bM}(V)\big)\,.
\end{equation}
The  real analytic inverse 
$\pi_{\cal E}:\ri\big({\bM}(V)\big)\to{\cal E}$ shall be called
{\it mean value parametrization}. Below we also write $\pi_{\cal E}$
for the map $\pi_{\cal E}\circ\pi_V$ defined on the
\emph{domain} ${\rm dom}\,\cE\;:=\;\st(\cA)\cap(\cE+V^{\perp})$, 
which was introduced in \cite{Ay} (for probability distributions).
In fact, the chart (\ref{eq:mean_chart}) was established in 
\cite{Wichmann} for $\cA={\rm Mat}(N,\mathbb C)$. Since $V$ 
contains only traceless matrices, its is proved in Lemma 3.13 in 
\cite{Weis_supp} that ${\bM}_{\cA}(V)={\bM}_{{\rm Mat}(N,\mathbb C)}(V)$ 
holds for every C*-subalgebra $\cA\subset{\rm Mat}(N,\mathbb C)$ which 
contains $V$. Remark~\ref{rem:supp_extend} extends this equality to
real *-subalgebras $\cA$ including $V$. So (\ref{eq:mean_chart}) holds
for these algebras.
\par
The two charts for a Gibbsian family $\mathcal E$ have open subsets of 
the canonical tangent space $V$ as their images. Given that $V$ is an 
affine space, two kinds of affine geodesics for $\cE$ arise: 
Unparametrized \emph{$(+1)$-geodesics} are the images of open segments
in $V$ under the canonical parametrization $\exp_1:V\to\cE$, and 
unparametrized \emph{$(-1)$-geodesics} are the images of open segments 
in $\ri\big({\bM}(V)\big)$ under the mean value parametrization 
$\pi_{\cE}:\ri\big({\bM}(V)\big)\to\cE$. 
We shall denote the \emph{open segment} between $a,b\in\sa$ 
by $]a,b[\,:=\{(1-\lambda)a+\lambda b\mid 0<\lambda<1\}$ and the 
\emph{closed segment}
by $[a,b]:=\{(1-\lambda)a+\lambda b\mid 0\leq\lambda\leq 1\}$.
A more comprehensive introduction of $(\pm)$-geodesics is given
in Section 7.2 in \cite{Amari}. The geodesics are part of a 
beautiful theory, called information geometry, about affine 
connections and Riemannian metrics on state spaces. See 
Remark~\ref{rem:details_geodesics} for some details.

\par
The relative entropy suits exponential families very well. If 
$\rho,\sigma$ and $\tau$ are states in $\cA$ with $\sigma$ and $\tau$ 
invertible, and if $\rho-\sigma\perp\ln(\tau)-\ln(\sigma)$, then 
\begin{equation}
\label{eq:pythagoras}\textstyle
S(\rho,\sigma)+S(\sigma,\tau)\;=\;S(\rho,\tau)
\end{equation}
holds, see e.g.\ \cite{PetzGeo}. This is the \emph{Pythagorean theorem} 
of the relative entropy. Clearly the Pythagorean theorem 
(\ref{eq:pythagoras}) holds if $\sigma$ and $\tau$ belong to an 
exponential family $\cE$ in $\cA$ and if $\rho\in\st(\cA)$ satisfies 
$\rho-\sigma\perp V$. The \emph{projection theorem} follows for 
$\rho\in{\rm dom}\,\cE$:
\begin{equation}
\label{eq:entropy_distance}\textstyle
\min_{\sigma\in\cE}S(\rho,\sigma)
\;=\;S\big(\rho,\pi_{\cE}(\rho)\big)\,,
\end{equation}
the minimum being unique. See Remark~\ref{rem:details_geodesics} 
about the information geometry of these theorems.
\par
The linearly constrained maximization of von Neumann entropy can be 
replaced by an unconstrained minimization of the relative entropy.
As mentioned previously, for $V=\Theta$ the mean value parametrization 
\begin{equation}
\label{eq:mean_par_inference}
\pi_{\cE}:\ri\big({\bM}(V)\big)\to\cE
\end{equation}
assigns to vectors $v\in\ri\big({\bM}(V)\big)$ the unique state 
$\pi_\cE(v)$
of maximum von Neumann entropy in the fiber 
$F(v):=(v+V^\perp)\cap\st(\cA)$. This is often proved using Lagrange
multipliers or Klein's inequality, see e.g.\ \cite{Ingarden}.
\par
A proof of (\ref{eq:mean_par_inference}) by information geometry 
opens a link to the entropy distance:
Let $\tau=\id/{\rm tr}(\id)$ be a reference state. 
Then $S(\rho,\tau)=-S(\rho)+\log({\rm tr}(\id))$ for all 
$\rho\in\st(\cA)$, so maximizing the von Neumann entropy is 
equivalent to minimizing the relative entropy from $\tau$, under
arbitrary constraints (a different choice of $\tau$ corresponds 
to a biased inference \cite{Ruskai}).
For all $v\in\ri\big({\bM}(V)\big)$ the fiber $F(v)$ is included in 
the domain  ${\rm dom}\,\cE$ of the Gibbsian family $\cE=\exp_1(V)$ 
by the mean value chart (\ref{eq:mean_chart}). Since $\tau\in\cE$, 
the Pythagorean theorem (\ref{eq:pythagoras}) shows for any state 
$\rho\in F(v)$
\[
S(\rho,\pi_\cE(\rho))+S(\pi_\cE(\rho),\tau)=S(\rho,\tau)\,.
\]
Minimizing $S(\cdot,\tau)$ over $\rho\in F(v)$ has the unique solution 
$\pi_\cE(\rho)$. In addition, by the projection theorem 
(\ref{eq:entropy_distance}), it is equivalent to the unconstrained
minimization of $S(\rho,\cdot)$ on $\cE$ (independent of the
choice of $\rho\in F(v)$).
\par 
Pythagorean and projection theorems as well as the $(+1)$- and 
$(-1)$-geodesics are rooted in information geometry.
\begin{Rem}
\label{rem:details_geodesics}
The exponential family $\ri\big(\st(\cA)\big)$ of invertible density
matrices has the mean value chart
$\rho\mapsto\rho-\id/{\rm tr}(\id)$. Its tangent space at $\rho$ is 
called the \emph{(m)-representation} 
and equals $\cA_0:=\{a\in{\cal A}_{\rm sa}\mid{\rm tr}(a)=0\}$, 
see p.\ 148 in \cite{Amari}. According to \cite{PetzGeo,Grasselli}, 
the BKM (Bogoliubov-Kubo-Mori) metric, a Riemannian metric on 
$\ri\big(\st(\cA)\big)$, can be defined for invertible density matrices 
$\rho$ and tangent vectors $A,B$ in the (m)-representation by
\[
g(A,B)_\rho
:=\int_0^\infty
{\rm tr}\left((t+\rho)^{-1}A(t+\rho)^{-1}B\right){\rm d}t\,.
\]
Although the BKM metric is a natural generalization of the Fisher metric 
to state spaces of 
{non-commutative}
algebras, unlike the former 
it is not the only such monotone one, see e.g.\ \cite{Petz3}.
\par
The \emph{(m)-connection} on the state space ${\rm ri}(\st(\cA))$,
denoted $\nabla^{\rm (m)}$, is defined through the parallel transport 
of translation on the affine hull 
${\cal A}_1=\{a\in{\cal A}_{\rm sa}\mid{\rm tr}(a)=1\}$ of the 
state space. If $g$ is a Riemannian metric on the manifold
${\rm ri}(\st(\cA))$ then the \emph{(e)-connection},
denoted $\nabla^{\rm (e)}$, is defined by
\[
X g(Y,Z)=g\big(\nabla^{\rm (m)}_XY,Z\big) + g\big(Y,\nabla^{\rm (e)}_XZ\big)
\]
for vector fields $X,Y,Z$ on $\ri\big(\st(\cA)\big)$. The connections
$\nabla^{\rm (m)}$ and $\nabla^{\rm (e)}$ are said to be \emph{dual} 
with respect to $g$. The (m)-connection is also called 
$(-1)$-connection, and when the BKM Riemannian metric $g$ is used, 
then the dual (e)-connection is called $(+1)$-connection. 
The connections $\nabla^{(+1)}$ and $\nabla^{(-1)}$ give rise to 
the geodesics introduced above, see e.g.\ Section 7.2 and
Section 7.3 in \cite{Amari}.
\par
The state space of the (real) *-subalgebra $\mathcal A$ is trivially
$(-1)$-autoparallel (i.e.\ totally geodesic) and it is $(+1)$-autoparallel 
as it is an exponential family. This shows that the $(\pm)$-connections
restrict from the state space of 
${\rm Mat}(N,\mathbb C)$ to ${\rm ri}(\mathcal S(\mathcal A))$.
\par
A Pythagorean theorem and a projection theorem are known in information 
geometry for \emph{dually flat spaces}. The relative entropy is the 
\emph{canonical divergence} of the dually flat space of 
invertible density matrices with respect to the BKM metric and 
the $(\pm)$-connections. Hence the Pythagorean theorem 
(\ref{eq:pythagoras}) arises from a more general theory, see 
e.g.\ Section 3.4 in \cite{Amari}. The $(-1)$-geodesic through $\rho$ 
and $\sigma$ and the $(+1)$-geodesic through $\sigma$ and $\tau$ meet 
at $\sigma$ orthogonally with respect to the BKM metric.
\end{Rem}
\par
The projection $\pi_V|_{\st(\cA)}:\st(\cA)\to V$ can be seen as a quantum 
channel to a commutative algebra.
\begin{Rem}
The mean value set $\bM(V)=\pi_V(\st(\cA))$ relates to a POVM quantum 
measurement.
A \emph{POVM} is defined as a finite sequence 
$F_1,\ldots,F_n$ of positive semidefinite matrices in $\cA$, such that 
$F_1+\cdots+F_n=\id$. The probability of outcome 
$i\in\{1,\ldots,n\}$ when measuring the quantum system $\rho\in\st(\cA)$ is
$\bP_\rho(i):=\tr(F_i\rho)$, see e.g.\ \cite{Petz}.
Given a POVM $F_1,\ldots,F_n$ in $\cA$, a quantum channel
\[
\st(\cA)\to\st(\mathbb C^n)\,,\quad
\rho\mapsto(\bP_\rho(1),\ldots,\bP_\rho(n))
=(\tr(F_1\rho),\ldots,\tr(F_n\rho))
\]
is defined. If $U$ is the real linear span of $F_1,\ldots,F_n$ and
$\widetilde{U}$ is the orthogonal projection of $U$ onto the space
of traceless matrices $\cA_0$, then the mean value sets
$\bM(U)=\pi_U(\st(\cA))$ and 
$\bM(\widetilde{U})=\pi_{\widetilde{U}}(\st(\cA))$ are affinely 
isomorphic to the image of the above channel
$\st(\cA)\to\st(\mathbb C^n)$. (For a proof see
e.g.\ Remark 1.1 in \cite{Weis_supp}).
\end{Rem}
\par
We would like to comment on (real) *-subalgebras.
\begin{Rem}
\label{rem:supp_extend}
As already mentioned earlier, *-subalgebras allow for 
low-dimensional examples. What makes *-subalgebra ${\cal A}$ of 
${\rm Mat}(N,\mathbb C)$ eligible for our treatment is that all 
results in \cite{Weis_supp} are true for them (unfortunately the 
choice in that article was to argue with intersections of 
C*-subalgebras and real matrices ${\rm Mat}(N,\mathbb R)$). Some
caution is 
needed, e.g.\ spectral projections of normal matrices need not 
be included in $\cA$, as the matrix ${\rm i}\sigma_3\oplus 0$ in 
Example~\ref{exa:algebra} shows. This error is present in 
Definition 2.5.3 of the above article. However, as only self-adjoint 
matrices are used, there is no problem arising. 
\par
An important feature of a *-subalgebra $\cA$ of 
${\rm Mat}(N,\mathbb C)$
is that spectral projections $p$ of a self-adjoint matrix 
$a\in{\cal A}_{\rm sa}$ can be written as $p=f(a)$ for a real 
polynomial $f$ in one variable. This implies that
\begin{itemize}
\item[$\bullet$]
if $a$ is a self-adjoint matrix and $g$ is a real valued function
defined on the spectrum of $a$, then $g(a)$ belongs to 
${\cal A}_{\rm sa}$,
\item[$\bullet$]
the state space has codimension one in ${\cal A}_{\rm sa}$, as the
cone of positive semi-definite matrices has full dimension
(decompose a self-adjoint matrix into a difference of two positive 
semi-definite matrices).
\end{itemize}
\par
One superficial flaw of *-subalgebras (and of C*-subalgebras!) is 
that eigenvalues can not be used directly, as the identity $\id$ of 
$\cA$ may differ from the identity $\id_N$ in ${\rm Mat}(N,\mathbb C)$.
On a closer inspection one realizes that this is exactly the 
flexibility we need
e.g.\ in Proposition~\ref{pro:inclusion} and
Theorem~\ref{thm:swallow} to analyze rI-closures.
The (+1)-closure of an exponential family is formed by exponential 
families of strictly smaller support, lying in 
\emph{compressed algebras} 
\[
p\cA p\;=\;\{pap\mid a\in\cA\}
\]
with identity $p=p^2=p^*\in\cA$. The algebra 
$p\cA p$ as a *-subalgebra of ${\rm Mat}(N,\mathbb C)$ may be treated
in the same way as $\cA$.
The unorthodox use of spectral values within a finite-dimensional
algebra was overlooked in \cite{Weis_supp}, see the correction
Lin.\ Alg.\ Appl.\ {\bf 436} no.\ 1 p.\ xvi (2012). 
\end{Rem}
%
%
%
%%%%%%%%%%%%%%%%%%%%%%%%%%%%%%%%%%%%%%%%%%%%%%%%%%%%%%%%%%%%%%%%%%
%%%%%%%%%%%%%%%%%%%%%%%%%%%%%%%%%%%%%%%%%%%%%%%%%%%%%%%%%%%%%%%%%%
%
%
\section{A classical---quantum metamorphosis}
\label{sec:cone_family}
\par
In the algebra $\cA$ from Example~\ref{exa:algebra} we study a curve 
of 2D mean value sets and we address the question whether 
they have non-exposed faces.
The algebra $\cA$ has the commutative *-subalgebra 
$\left(\begin{smallmatrix}*&0&0\\{}0&*&0\\0&0&*\end{smallmatrix}\right)$
of diagonal matrices, isomorphic to $\mathbb R^3$, and its 
left upper corner 
$\left(\begin{smallmatrix}*&*&0\\{}*&*&0\\0&0&0\end{smallmatrix}\right)$
is a non-commutative *-subalgebra. 
\par
The curve of mean value sets is rather a Grassmannian manifold of 
subspaces. More precisely, we consider 2D subspaces of the 4D 
space $\sa$ of self-adjoint matrices and here we restrict to 2D
subspaces of the 3D space of traceless matrices (since the state 
space is parallel to it). So by symmetry of the cone $\mathcal S(\mathcal A)$
one real angular variable suffices to describe mean value sets. Thus
we can consider a curve in the Grassmannian manifold.
In Figure~\ref{fig:reflections}, left, mean value sets 
$\bM(V)$ are drawn isometrically at equidistant $\tfrac 1{12}\pi$ 
angles around a full circle.
\par
Our example is minimal in two respects:
\begin{itemize}
\item[$\bullet$]
Planar projections have minimal dimension to allow for non-exposed 
faces.
\item[$\bullet$]
The algebra $\mathcal A$ is (up to isomorphism)
the smallest *-subalgebra $\cA$ of 
${\rm Mat}(N,\mathbb C)$ allowing for a mean value set with non-exposed 
faces. If $\cA$ has no *-subalgebra isomorphic to ${\rm Mat}(2,\mathbb R)$ 
then, assuming $\id=\id_N$, then $\cA$ is commutative
(see Theorem 5.2 and 5.4 in Section 5 in \cite{Kojima}).
($\id=\id_{\widetilde N}$ may be achieved by restricting a faithful 
representation of the C*-algebra $\id\,{\rm Mat}(N,\mathbb C)\id$ onto a 
direct sum of full matrix algebras, see e.g.\ \cite{Davidson}.)
Hence the state space $\mathcal S(\mathcal A)$ is a simplex. Then all mean 
value sets are polytopes having no non-exposed faces. The algebra 
${\rm Mat}(2,\mathbb R)
\cong{\rm span}_\bR\{\id_2,\sigma_1,\sigma_2,{\rm i}\sigma_3\}$ itself has 
a disk as state space, whose 
proper projections are a point or a segment, having no 
non-exposed faces.
\end{itemize}
\par
We introduce precise definitions in convex geometry 
for subsequent discussions e.g.\ in 
Lemma~\ref{lemma:infimum_at_infty}.
\begin{Def}
Let $M$ be a compact and convex subset of a finite-dimensional
Euclidean vector space $(\bE,\langle\cdot,\cdot\rangle)$. 
\begin{enumerate}
\item[$\bullet$]
A convex 
subset $F$ of $M$ is a \emph{face} of $M$, if for all $x,y\in M$ and all 
$0<\lambda<1$ the inclusion of $(1-\lambda)x+\lambda y\in F$ implies 
$x,y\in F$. 
\item[$\bullet$]
A face of dimension zero is called \emph{extreme point} and if it is
not exposed, a \emph{non-exposed point}.
An extreme point of $\st(\cA)$ will be called \emph{pure state}.
\item[$\bullet$]
If $M$ is non-empty, then for non-zero $u\in\bE$ 
the \emph{supporting hyperplane} is defined by
\[
H(M,u)\;:=\;
\{x\in\bE\mid\langle x,u\rangle=\max_{y\in M}\langle y,u\rangle\}\,.
\]
\item[$\bullet$]
A face $F$ of $M$ is called
\emph{exposed} if $F$ is the intersection of $M$ with a 
supporting hyperplane 
\begin{equation}
\label{eq:exposed_face}
F(M,u)\;:=\;M\cap H(M,u)\,.
\end{equation}
$F=\emptyset$ and $F=M$ are exposed faces by definition.
\end{enumerate}
\end{Def}
\par
The Grassmannian manifold $\bG$ of real 2D subspaces of
self-adjoint traceless matrices $\cA_0=\{a\in\sa\mid\tr(a)=0\}$
will be denoted
\[
\bG \; := \;
\{V\subset\cA_0\mid V\text{ is a real 2D subspace }\}\,.
\]
We define the angle between a subspace $V\in\bG$ and 
$z:=(-\frac{1}{2}\id_2)\oplus 1$ 
(pointing along the axis of the cone),

\begin{equation}
\label{eq:phi}
\varphi \; = \; \varphi(V) \; := \; \winkel(V,z)\,.
\end{equation}
\par
The state space is $\st(\cA)={\rm conv}(\st(\cB)\cup\{0_2\oplus 1\})$
for the disk $\st(\cB)$ introduced in Example~\ref{exa:algebra}.
The mean value set of $V\in\bG$ is the convex hull of the ellipse 
$e:=\pi_V(\st(\cB))$ and of $x:=\pi_V(0_2\oplus 1)$,
\begin{equation}
\label{eq:mean_set}
\bM(V)\;=\;{\rm conv}(e,x)\,.
\end{equation}
The problem of finding non-exposed faces at $\bM(V)$ may be solved
in $\bR^3$ by studying projections of a symmetric 3D cone isometric to
$\cS(\cA)$. Explicit calculations
 with matrices are done in Example 1.2 in 
\cite{Weis_supp} by studying tangents to the elliptical boundary curve 
$\partial e$. For all subspaces $V\in\bG$ the projection of $V$ onto 
${\rm span}_\bR(\sigma_1,\sigma_2,\sigma_3)\oplus 0$ is a 
subspace of ${\rm span}_\bR(\sigma_1,\sigma_2)\oplus 0$. Hence the state 
space $\st(\cA)$ equals the cone $C$ in \cite{Weis_supp} and we have
the following:
\begin{Lem}
\label{Lem:nonex}
Let $V\in\bG$ be a 2D plane. If $\varphi=0$, then $\partial e$ is 
a segment (degenerate ellipse) and the mean value set $\bM(V)$ is 
a triangle. If $0<\varphi<\frac{\pi}{3}$, then $\partial e$ is a 
non-degenerate ellipse, $x\not\in e$ and the tangents from $x$ to 
$\partial e$ meet $\partial e$ at two non-exposed points of $\bM(V)$. 
If $\frac{\pi}{3}\leq\varphi\leq\frac{\pi}{2}$, then 
$\bM(V)=e$ is bounded by a non-degenerate ellipse $\partial e$.
\end{Lem}
\par
We see that non-exposed faces are typical in the following 
sense.
A continuous curve $\gamma:$ $[0,1]\to\bG$ induces a curve of mean value 
sets $\lambda\mapsto\bM(\gamma(\lambda))$. By Lemma~\ref{Lem:nonex} 
a mean value set without non-exposed faces must be a triangle or an 
ellipse. If $\gamma$ connects the classical mean value set of a triangle 
to an ellipse, then we have $\winkel(\gamma(0),z)=0$ and
$\winkel(\gamma(1),z)\in[\frac{\pi}{3},\frac{\pi}{2}]$. Since the angle 
$\varphi$ is continuous on $\bG$, the curve $\gamma$ must cross the range 
of angles $(0,\frac{\pi}{3})$ with mean value sets having non-exposed 
faces. This range corresponds to an open subset of the Grassmannian 
$\bG$. 
%
%
%
%%%%%%%%%%%%%%%%%%%%%%%%%%%%%%%%%%%%%%%%%%%%%%%%%%%%%%%%%%%%%%%%%%%%%%%%
%%%%%%%%%%%%%%%%%%%%%%%%%%%%%%%%%%%%%%%%%%%%%%%%%%%%%%%%%%%%%%%%%%%%%%%%
%%%%%%%%%%%%%%%%%%%%%%%%%%%%%%%%%%%%%%%%%%%%%%%%%%%%%%%%%%%%%%%%%%%%%%%%
%
%
\section{Closures of exponential families}
\label{sec:closures}
\par
The curve of 2D mean value sets $\bM(V)$ in Section~\ref{sec:cone_family} 
shows that the angle of $\varphi=\varphi(V)=\frac{\pi}{3}$ divides mean 
value sets with non-exposed faces from others without non-exposed faces. 
In Section~\ref{sec:staffelberg} we show that the Gibbsian family at 
$\varphi=\frac{\pi}{3}$, called Staffelberg family, has a discontinuous 
entropy distance. The analysis is based on more general results about 
$(+1)$-closures in Section~\ref{sec:general}. In
Section~\ref{sec:-1-staffelberg} we compute the $(-1)$-closure
of the Staffelberg family. We will see in Section~\ref{sec:swallow} that
the $(+1)$-closure of a Gibbsian family, in general, is not a locus of 
maximum-entropy density matrices under linear constraints.
\par
In the sequel we assume that $\cA$ is a real *-subalgebra of 
${\rm Mat}(N,\bC)$ and that $\cE$ is an exponential family in $\cA$
with canonical parameter space $\Theta$ and canonical tangent space 
$V=\lin(\Theta)$.
In Section~\ref{sec:staffelberg} through~\ref{sec:swallow} we shall
specialize to the algebra $\cA$ defined in Example~\ref{exa:algebra}.

%
%%%%%%%%%%%%%%%%%%%%%%%%%%%%%%%%%%%%%%%%%%%%%%%%%%%%%%%%%%%%%%%%%%%%%%
%%%%%%%%%%%%%%%%%%%%%%%%%%%%%%%%%%%%%%%%%%%%%%%%%%%%%%%%%%%%%%%%%%%%%%
%%%%%%%%%%%%%%%%%%%%%%%%%%%%%%%%%%%%%%%%%%%%%%%%%%%%%%%%%%%%%%%%%%%%%%
%
%
\subsection{$(+1)$-closures of exponential families}
\label{sec:general}
\par
In this section we compute the $(+1)$-closure ${\rm cl}^{(+1)}(\cE)$ 
defined in (\ref{eg:1_closure}). We show that it is a union of 
exponential families. We also discuss aspects of the rI-closure 
${\rm cl}^{\rm rI}({\cal E})$, defined in (\ref{eg:rI_closure})
and of the norm closure $\overline{\cE}$. Among others, we show
\[\textstyle
{\rm cl}^{(+1)}(\cE)
\;\subset\;
{\rm cl}^{\rm rI}(\cE)
\;\subset\;
\overline{\cE}\,.
\]
Strict inclusions are presented by example in 
Section~\ref{sec:swallow} and Section~\ref{sec:staffelberg}
\par
In this section $\cA$ denotes an arbitrary real *-subalgebra
of ${\rm Mat}(N,\mathbb C)$.
In the analysis of $(+1)$- and rI-closures, subalgebras 
with various identities will appear, so spectral values shall be used
in some statements, see also Remark~\ref{rem:supp_extend}.
On the space $\sa$ of self-adjoint matrices
we have the partial ordering defined by $a\preceq b$ 
if and only if $b-a\succeq 0$, i.e.\ $b-a$ is positive semi-definite.
The set of projections $\{p\in\cA\mid p^*=p^2=p\}$ will be considered
with this partial ordering. If $p\in\cA$ is a projection, then the 
\emph{compressed algebra} by $p$ is
\begin{equation}
\label{eq:compressed_algebra}
p\cA p\;:=\;\{p\,ap\mid a\in\cA\}\,.
\end{equation}
The algebra $p\cA p$ is a *-subalgebra of $\cA$ with identity $p$. 
The spectral values of $a\in\sa$ are the real numbers $\lambda$ such 
that $a-\lambda\id$ is not 
invertible in $\cA$. The sum of spectral projections of non-zero 
spectral values of $a$ is the \emph{support projection} $s(a)$;
we notice 
\begin{equation}
\label{eq:support_projection}
s(a)\in\cA\,.
\end{equation} 
We denote by $\lambda^+(a)$ the maximal 
spectral value of $a$ and by $p^+(a)\in\cA$ the spectral projection 
of $a$ corresponding to $\lambda^+(a)$, which we call the 
\emph{maximal projection} of $a$. 
Notice in Remark~\ref{rem:supp_extend} that eigenvalues can not be 
used.
\par
The \emph{free energy}, defined for $a\in\sa$ by $F(a):=\ln(\tr(e^a))$
is useful to discuss limits of $(+1)$-geodesics. Functions defined
for projections $p\in\cA$ by functional calculus on $(p\cA p)_{\rm sa}$ 
will be decorated by a superscript $p$, e.g.\ $\ln^p(p)=0$, while 
$\ln(p)$ is not defined if $p\neq\id$. The superscript  $p=\id$
will often be omitted.
For $a\in(p\cA p)_{\rm sa}$ we notice 
$\exp^p(a)=p\exp(a)$, $\exp_1^p(a)=\frac{p\,e^a}{\tr(p\,e^a)}$ and 
$F^p(a)=\ln\tr(p\,e^a)$. We use the projection
$\cA\rightarrow p\cA p$, $a\mapsto pap$ to define the exponential 
family in $p\cA p$
\[
\cE^p \; := \; 
\left\{\exp_1^p( p\theta p )\mid\theta\in\Theta\right\}\,.
\]
\begin{Lem}
\label{lem:e-limits}
Suppose $\theta,u\in\sa$ and $p:=p^+(u)$ is the maximal projection of $u$. 
We have
\begin{equation}
\label{asy:exp1}
\lim_{t\to\infty}\exp_1(\theta+t\,u)
\;=\;{\exp_1^p(p\theta p)}
\end{equation}
and
\begin{equation}
\label{asy:free_energy}
\lim_{t\to\infty}\big(F(\theta+t\,u)-t\,\lambda^+(u)\big)
\;=\;{F^p(p\theta p)\,.}
\end{equation}
\end{Lem}
{\em Proof:\/} 
If $u$ has maximal 
{spectral value}
$\lambda^+(u)=0$ then by standard perturbation theory one proves
\begin{equation}
\label{eq:boundary}
\lim_{t\to\infty}e^{\theta+t\,u}
\;=\;
p\,e^{p\,\theta p}\,.
\end{equation}
Since $\exp_1(\theta+\alpha\id)=\exp_1(\theta)$ holds for 
$\alpha\in\bR$ we have for arbitrary $u\in\sa$
\[\textstyle
\lim_{t\to\infty}\exp_1(\theta+t\,u)
\;=\;
\lim_{t\to\infty}\exp_1(\theta+t\,(u-\lambda^+(u)\id))
\;=\;
\frac{p\,e^{p\,\theta p}}{\tr(p\,e^{p\,\theta p})}\,.
\]
If $u$ has maximal
{spectral value}
$\lambda^+(u)=0$ then
(\ref{eq:boundary}) and the continuity of the logarithm show
$\lim_{t\to\infty}F(\theta+t\,u)=\ln\tr(p\,e^{p\,\theta p})$.
We have $F(\theta+\alpha\id)=F(\theta)+\alpha$ for 
$\alpha\in\bR$, hence for arbitrary $u\in\sa$ the equality of
\[
F(\theta+t\,u)-t\lambda^+(u)
\;=\;
F[\theta+t\,(u-\lambda^+(u))]
\]
shows the second claim.
\hspace*{\fill}$\Box$\\
\par
An immediate consequence of (\ref{asy:exp1}) is as follows.
\begin{Pro}
\label{pro:geodesic_closure}
The $(+1)$-closure of $\cE$ is ${\rm cl}^{(+1)}(\cE)=\bigcup_p\;\cE^p$
where the disjoint union extends over the maximal projections $p=p^+(v)$ 
of all vectors $v\in V$ (including $\id=p^+(0)$). 
\end{Pro}
\par
The first hurdle to tackle the rI-closure will be
Lemma~\ref{lemma:infimum_at_infty} which controls limits of
the relative entropy of certain states $\rho$ from states 
$\sigma$ on $(+1)$-geodesics. This is remarkable since for 
$\cA={\rm Mat}(2,\bC)$
\[\textstyle
S_{\rho}(\sigma)\;:=\;S(\rho,\sigma)
\]
is not continuous on the set 
$\{\sigma\in\st\mid s(\sigma)\succeq s(\rho)\}$
with larger support projections (\ref{eq:support_projection}). 
However, $S_{\rho}$ is continuous throughout
the simplex $\st$ for $\cA\cong\bC^N$.
\begin{Exa}[Discontinuity of Relative Entropy]
In the algebra $\cA={\rm Mat}(2,\bC)$ of a qubit
we consider the pure state $\rho:=\tfrac 1{2}(\id_2+\sigma_1)$.\\
For real $\alpha>0$ let $s_\alpha\in[0,1]$ such that 
$s_\alpha\stackrel{\alpha\to 0}{\to}0$ and define the state
\[\textstyle
\sigma_{\alpha}
\;:=\;
(1-s_\alpha)\tfrac 1{2}
(\id_2+\cos(\alpha)\sigma_1+\sin(\alpha)\sigma_2)
+s_\alpha\,\tfrac 1{2}(\id_2-\cos(\alpha)\sigma_1-\sin(\alpha)\sigma_2)\,.
\]
Then $\sigma_\alpha\stackrel{\alpha\to 0}{\to}\rho$ as well as
\[\textstyle
S(\rho,\sigma_\alpha)
\;=\;
-\tfrac 1{4}\alpha^2\log(s_\alpha)(1+o(1))+o(1)\,.
\]
E.g.\ if we choose $c,\gamma>0$ and put 
$s_\alpha:=\exp(-c/\alpha^\gamma)$, then
$s_\alpha\stackrel{\alpha\to 0}{\to}0$ and
\[\textstyle
S(\rho,\sigma_\alpha)
\;=\;
\tfrac c{4}\alpha^{2-\gamma}(1+o(1))+o(1)\,.
\]
So any non-negative limit of $S(\rho,\sigma_\alpha)$ can be achieved
for smooth paths converging to an arbitrary point $\rho$ in the
boundary of state space.
\end{Exa}
\par
Using maximal spectral values $\lambda^+$ and maximal projections
$p^+$ we summarize Proposition 2.9 in \cite{Weis_supp}.
\begin{Lem}
\label{lem:exposed_of_state}
If $u\in\sa$ is a non-zero self-adjoint matrix, then the exposed face 
$F(\st(\cA),u)$ consists of the states $\rho\in\st(\cA)$ such that 
$\langle \rho,u\rangle=\lambda^+(u)$ or, equivalently, 
$s(\rho)\preceq p^+(u)$. 
\end{Lem}
\par
The lemma says that the exposed face $F(\st(\cA),u)$ is the state 
space of the compressed algebra $p\cA p$ discussed in 
(\ref{eq:compressed_algebra}) for $p:=p^+(u)$. Moreover, it follows 
that all faces of $\st(\cA)$ are exposed, see e.g.\ Section 2.3 in 
\cite{Weis_supp}.
\par
The derivative of the exponential function for $a,b\in\sa$ is
\[
{\rm D}|_a\exp(b)
\;=\;
\int_0^1e^{ya}be^{(1-y)a}{\rm d}y\,.
\]
It implies the derivative of the free 
energy $F$
\begin{equation}\textstyle
\label{eq:free_derivative}
{\rm D}|_a F(b)\;=\;\langle b,\exp_1(a)\rangle\,.
\end{equation}
The derivative of the exponential for $\cA={\rm Mat}(N,\bC)$
is explained by power series expansion e.g.\ in \cite{Lieb} 
and may be generalized to any *-subalgebra 
$\cA$ of ${\rm Mat}(N,\bC)$ by left- and right-multiplication 
with the identity $\id$ in $\cA$.
\begin{Lem}
\label{lemma:infimum_at_infty}
Suppose $\theta,u\in\sa$ such that $u$ is not a multiple of the 
identity $\id$ in $\cA$ and let $p:=p^+(u)$. If
$\rho\in F(\st(\cA),u)$, 
then $S_{\rho}(\exp_1(\theta+t\,u))$ is strictly monotone decreasing 
with $t\in\bR$ and
\[
S_{\rho}\big(\exp_1^p(p\,\theta p)\big)
\;=\;
\lim_{t\to\infty}S_{\rho}\big(\exp_1(\theta+t\,u)\big)
\;=\;
\inf_{t\in\bR}S_{\rho}\big(\exp_1(\theta+t\,u)\big)\,.
\]
\end{Lem}
{\em Proof:\/} 
By definition (\ref{eq:exposed_face}) of an exposed face we have for 
$\tau\in\st(\cA)$ and for $\rho\in F(\st(\cA),u)$ the inequality
$\langle u,\tau-\rho\rangle\leq 0$. Since $u$ is not proportional
to $\id$, its maximal projection is $p:=p^+(u)$ is not $\id$.
If $\tau$ is invertible, then $s(\tau)=\id$ and it follows from
Lemma~\ref{lem:exposed_of_state} that $\tau\not\in F(\st(\cA),u)$. 
This implies the strict inequality $\langle u,\tau-\rho\rangle\;<\;0$
to hold for all invertible states $\tau=\exp_1(\theta+t u)$ 
with $t\in\bR$. Using (\ref{eq:free_derivative}) we have for
all $t\in\bR$
\[\textstyle
\frac{\partial}{\partial\lambda}
S_{\rho}\circ\exp_1(\theta+t\,u)
\;=\;\langle u,\exp_1(\theta+t\,u)-\rho\rangle
\;<\;0\,.
\]
We conclude that $S_{\rho}\circ\exp_1(\theta+t\,u)$ is strictly 
monotone decreasing in $t$\,.
\par
The limit of the $(+1)$-geodesic
$g:t\mapsto\exp_1(\theta+t\,u)$ is calculated in (\ref{asy:exp1}),
\[\textstyle
\sigma\;:=\;\lim_{t\to\infty}g(t)
\;=\;\exp_1^p(p\,\theta p)\,.
\]
The states $\rho$ and $\sigma$ belong to the compressed algebra $p\cA p$ 
defined in (\ref{eq:compressed_algebra}) and $\sigma$ is invertible in 
$p\cA p$. Then
\begin{eqnarray*}
\lefteqn{\textstyle
-S(\rho,\sigma)-S(\rho)\;=\;\tr\left(\rho\,\ln^p\circ
\exp_1^p(p\,\theta p)\right)
\;=\;\tr(\rho\,\theta)-F^p(p\,\theta p)}\\
& = & \textstyle
\lim_{t\to\infty}\left[\tr(\rho\,\theta)+t\,\lambda^+(u)
-F(\theta+t\,u)\right]\\
& = & \textstyle
\lim_{t\to\infty}\left[\tr\big(\rho\,(\theta+t\,u)\big)
-F(\theta+t\,u)\right]\\
& = & \textstyle
\lim_{t\to\infty}
\tr\big(\rho\,\ln\circ\exp_1(\theta+t\,u)\big)
\;=\; \lim_{t\to\infty}\left[-S(\rho,g(t))-S(\rho)\right]\,.
\end{eqnarray*}
We have used (\ref{asy:free_energy}) in the third step. The 
result is 
$\lim_{t\to\infty}S_{\rho}\circ g(t)=S_{\rho}(\sigma)$. 
Since $S_{\rho}\circ g$ is monotone decreasing in $\lambda$ we have 
$\inf_{t\in\bR}S_{\rho}\circ g(t)=S_{\rho}(\sigma)$.
\hspace*{\fill}$\Box$\\
\par
We show that $(+1)$-closures do not decrease the entropy distance, 
defined in~(\ref{eq:entropy_distance_def}), from exponential 
families.
\begin{Pro}
\label{pro:inclusion}
If $v\neq 0$ belongs to the canonical tangent space $V$ of the
exponential family $\cE$ and $\rho$ to the exposed face 
$F(\st(\cA),v)$, then 
${\rm d}_\cE(\rho)={\rm d}_{\cE^{p^+(v)}}(\rho)$. For arbitrary 
$\rho\in\st(\cA)$ we have ${\rm d}_\cE(\rho)
=\inf\{S(\rho,\sigma)\mid\sigma\in{\rm cl}^{(+1)}(\cE)\}$.
\end{Pro}
{\em Proof:\/}
We prove the first statement, let $p:=p^+(v)$. If $p^+(v)=\id$, then 
there is nothing to prove. Otherwise we have 
by Lemma~\ref{lemma:infimum_at_infty} and 
Lemma~\ref{lem:e-limits}
\begin{eqnarray*}
\lefteqn{
{\rm d}_\cE(\rho)
\;=\;\inf_{\sigma\in\cE}S(\rho,\sigma)
\;=\;\inf_{\theta\in\Theta}\inf_{t\in\bR}
S(\rho,\exp_1(\theta+t v))}\\
& = & \inf_{\theta\in\Theta}S(\rho,\lim_{t\to\infty}
\exp_1(\theta+t v))
\;=\; \inf_{\theta\in\Theta}S(\rho,\exp_1^p(p\theta p))
\;=\; {\rm d}_{\cE^p}(\rho)\,.\hspace{0.0cm}
\end{eqnarray*}
\par
For the second statement, let $\rho\in\st(\cA)$ be arbitrary. By
Proposition~\ref{pro:geodesic_closure} it suffices to show 
${\rm d}_{\cE^p}(\rho)\geq {\rm d}_{\cE}(\rho)$ for all projections $p$
of the form $p=p^+(v)$ where $v\in V$ is non-zero. If
$\rho\not\in F(\st(\cA),v)$, then $s(\rho)\not\preceq p$ by
Lemma~\ref{lem:exposed_of_state}. So for all
$\sigma\in\cE^p$ we have $S(\rho,\sigma)=\infty$. Otherwise, the equality
${\rm d}_{\cE^p}(\rho)={\rm d}_{\cE}(\rho)$ follows from the first
assertion above.
\hspace*{\fill}$\Box$\\
\begin{Cor}
\label{cor:inclusions}
We have ${\rm cl}^{(+1)}(\cE)
\subset{\rm cl}^{\rm rI}(\cE)\subset\overline{\cE}$. 
\end{Cor}
{\em Proof:\/}
The first inclusion follows from Proposition~\ref{pro:inclusion}:
If $\rho\in{\rm cl}^{(+1)}(\cE)$, then ${\rm d}_\cE(\rho)
=\inf\{S(\rho,\sigma)\mid\sigma\in{\rm cl}^{(+1)}(\cE)\}=0$
shows $\rho\in{\rm cl}^{\rm rI}(\cE)$.
\par
The second inclusion follows from the Pinsker-Csisz\'ar 
inequality (see e.g.\ p.\ 40 in \cite{Petz}), which says that
$\|\rho-\sigma\|_1^2\leq\frac{1}{2}S(\rho,\sigma)$ holds for all states 
$\rho,\sigma\in\st(\cA)$ with the trace norm
$\|a\|_1:=\tr(\sqrt{a^*a})$ for $a\in\cA$.
\hspace*{\fill}$\Box$\\
\par
Finally we prove an upper bound for the norm closure of a Gibbsian 
family.
\begin{Lem}
\label{lem:upper_bound}
Let $\cE$ be a Gibbsian family, i.e.\ $\Theta=V$. Then
$\overline{\cE}\subset\cE\cup\bigcup_v F\big(\st(\cA),v\big)$ where the union 
of exposed faces extends over all non-zero vectors $v\in V$.
\end{Lem}
{\em Proof:\/}
We assume $\theta_i\subset\Theta$ and that
$\rho_i:=\exp_1(\theta_i)\in\cE$ is a converging sequence with limit
$\rho:=\lim_{i\to\infty}\rho_i$. If $\pi_V(\rho)\in{\rm ri}(\bM(V))$,
then there is a neighborhood $U(\pi_V(\rho))\subset{\rm ri}(\bM(V))$
containing $\pi_V(\rho_i)$ for large $i$.
Choosing this neighborhood sufficiently small we can assume its
closure $X$ is a compact subset of ${\rm ri}(\bM(V))$. As discussed 
in (\ref{eq:mean_chart}) the map 
$\pi_V\circ\exp_1:V\to{\rm ri}(\bM(V))$ is a real analytic 
diffeomorphism. Using the inverse mapping, the set 
$\log_0\circ\pi_\cE(X)\subset V$ is compact and it contains
$\theta_i$ for large $i$. It follows $\rho\in\cE$.
\par
Otherwise, if $\pi_V(\rho)$ belongs to the boundary of the mean value
set, then by Theorem 13.1 in \cite{Rockafellar} there is a non-zero 
vector $v\in V$ such that $\pi_V(\rho)\in F(\bM(V),v)$. Then the state
$\rho$ lies in the exposed face $F(\st(\cA),v)$ for the same vector $v$.
\hspace*{\fill}$\Box$\\
%
%%%%%%%%%%%%%%%%%%%%%%%%%%%%%%%%%%%%%%%%%%%%%%%%%%%%%%%%%%%%%%%%%%%%%%%%
%%%%%%%%%%%%%%%%%%%%%%%%%%%%%%%%%%%%%%%%%%%%%%%%%%%%%%%%%%%%%%%%%%%%%%%%
%%%%%%%%%%%%%%%%%%%%%%%%%%%%%%%%%%%%%%%%%%%%%%%%%%%%%%%%%%%%%%%%%%%%%%%%
%
%
\subsection{The Staffelberg family}
\label{sec:staffelberg}
\par
The exponential family $\cE$ discussed in this section 
is an example of a discontinuous maximum-entropy 
inference announced in the introduction.
That exponential family has the form of the {\em Staffelberg} table 
mountain, in the natural preserve of 
{\em Fr\"ankische Schweiz---Veldensteiner Forst}.
Its mean value set appears at the angle (\ref{eq:phi}) of 
$\varphi=\frac{\pi}{3}$ in the metamorphosis of 
Figure~\ref{fig:reflections}. 
Smaller angles
$\varphi$ have non-exposed faces, larger angles do not. We explain the 
geometrical components of the closures
${\rm cl}^{(+1)}(\cE)={\rm cl}^{\rm rI}(\cE)\subsetneq\overline{\cE}$.
Then we address continuity issues. 
\begin{figure}[t!]
\begin{center}
\begin{picture}(12,6)
\ifthenelse{\equal{\eps}{ja}}{%
\put(0,0){\includegraphics[height=6cm, bb=50 30 500 500, clip=]%
{Fig41.eps}}
\put(5,0){\includegraphics[height=6cm, bb=0 20 500 373, clip=]%
{Fig42.eps}}}{%
\put(0,0){\includegraphics[height=6cm, bb=50 30 500 500, clip=]%
{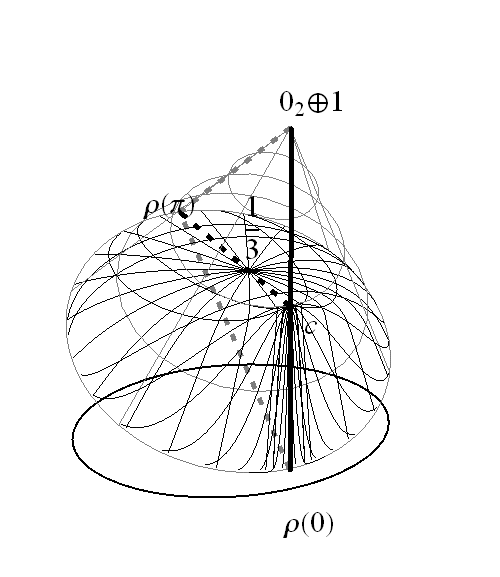}}
\put(5,0){\includegraphics[height=6cm, bb=0 20 500 373, clip=]%
{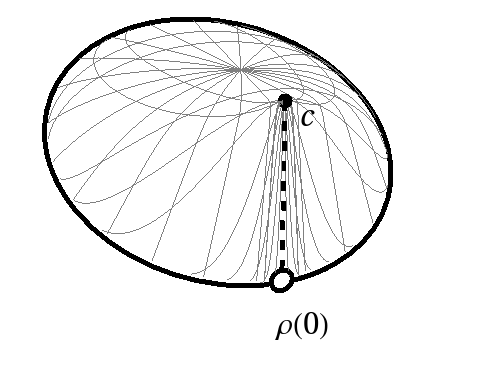}}}
%\put(0,0){\line(1,0){6}}
%\put(0,6){\line(1,0){6}}
%\put(0,0){\line(0,1){6}}
%\put(6,0){\line(0,1){6}}
%
%\put(6,0){\line(1,0){6}}
%\put(6,6){\line(1,0){6}}
%\put(6,0){\line(0,1){6}}
%\put(12,0){\line(0,1){6}}
\end{picture}
\end{center}
\caption{\label{fig:staffelberg}
The Staffelberg family $\cE$ sketched by $(+1)$-geodesics. Left: The 
cone about $\cE$ is the state space $\st(\cA)$. The ellipse below is 
the boundary of the mean 
value set $\bM(V)$. The generating line $[\rho(0),0_2\oplus 1]$ 
of the cone $\st(\cA)$, with midpoint $c$, is perpendicular to $V$. 
Right: $\cE$ has equal $(+1)$- and rI-closures, they cover 
the punctured base circle of $\st(\cA)$ (large circle) with $\rho(0)$
missing (small circle). These closures include $c$. The norm closure
of $\cE$ contains in addition the entire segment $[\rho(0),c]$.}
\end{figure}
\begin{Def}
\label{def:Staffelberg}
The \emph{Staffelberg family}, depicted in Figure~\ref{fig:staffelberg}, 
is the Gibbsian family
\[
\cE\;:=\;\exp_1\left({\rm span}_\bR\{\sigma_1\oplus 0,
\sigma_2\oplus 1\}\right)
\]
in the *-subalgebra $\cA\subset{\rm Mat}(3,{\mathbb C})$ defined in 
Example~\ref{exa:algebra}.
\end{Def}
\par
The self-adjoint matrices in $\cA$ are 
$\sa={\rm span}_{\bR}\{\id_2\oplus 0,\sigma_1\oplus 0,\sigma_2\oplus 0,
0_2\oplus 1\}$, the state space $\st(\cA)$ is a 3D cone. We use the 
notation 
\[
B:=\{\rho(\alpha)\mid\alpha\in(0,2\pi)\}
\]
for the \emph{punctured base circle} of $\st(\cA)$ with 
$\rho(0)=\frac{1}{2}(\id_2+\sigma_2)\oplus 0$ missing. The symmetry axis 
$l$ of $\st(\cA)$ goes through the tracial state $\frac{1}{3}\id$ and 
through the apex $0_2\oplus 1$, where it meets the generating lines of 
the cone $\st(\cA)$ under an angle of $\frac{\pi}{6}$. The generating 
line $[\rho(0),0_2\oplus 1]$ is perpendicular to $V$. We denote its 
midpoint by 
\[
c:=\tfrac{1}{2}(\rho(0)+0_2\oplus 1)\,.
\]
The canonical tangent space $V=\Theta$ of $\cE$ is spanned by 
$v_1:=\sigma_1\oplus 0$ and $v_2:=\sigma_2\oplus 1-\frac{1}{3}\id$. The 
vector $z=-\frac{1}{2}\id_2\oplus 1$ is perpendicular to $v_1$, so 
\[\textstyle
\varphi\;=\;\winkel(V,z)\;=\;\winkel(v_2,z)
\;=\;{\rm arccos}(\frac{1}{2})\;=\;\frac{\pi}{3}
\]
as claimed. The basis vectors of $V$ connect special points in $\st(\cA)$, 
\[\textstyle
v_1
\;=\;
\rho(\frac{\pi}{2})-\rho(\frac{3}{2}\pi)
\quad\textrm{and}\quad
v_2
\;=\;
\frac{4}{3}\left(c-\rho(\pi)\right)\,.
\]
\par
The *-algebra generated by $\sigma_2\oplus 1$ is isomorphic to $\bR^2$ 
and it has the segment $[\rho(\pi),c]$ as its state space. The 
$(+1)$-geodesic $\{\exp_1(\lambda v_2)\mid\lambda\in\bR\}$ is included 
in $\cE$ and it covers the invertible states in $[\rho(\pi),c]$. 
The *-algebra generated by $\rho(0)$, $\rho(\pi)$ and $0_2\oplus 1$
is isomorphic to $\bR^3$, its state space is the equilateral triangle
spanned by these generators, see Figure~\ref{fig:staffelberg}, left.
\par
For discussions of $(+1)$-geodesics in $\cE$ we use a redundant 
parametrization and define for real $\alpha,s,t$
\begin{align}
\label{eq:def_polar_ste}
E(\alpha,s,t)
\;:=\;  
\exp_1\big\{ & t\,
[\cos(\alpha)(\sigma_2\oplus 1)+\sin(\alpha)\sigma_1\oplus 0] \\\nonumber
 & +s\,
[-\sin(\alpha)(\sigma_2\oplus 1)+\cos(\alpha)\sigma_1\oplus 0]
\big\}\,.
\end{align}
Let $x:=s\cos(\alpha)+t\sin(\alpha)$, $y:=-s\sin(\alpha)+t\cos(\alpha)$,
$b:=\sqrt{x^2+y^2}=\sqrt{s^2+t^2}$ and
$\eta:=2\cosh(b)+e^{-s\sin(\alpha)+t\cos(\alpha)}$. Then
\begin{align*}
E(\alpha,s,t)
\;=\;  
\tfrac 1{\eta}\big\{ 
[\cosh(b)\id_2+\sinh(b)(x\sigma_1+y\sigma_2)/b]
\oplus e^{-s\sin(\alpha)+t\cos(\alpha)}\big\}\,.
\end{align*}
The vectors $v_1$ and $v_2$ are completed by $v_3:=0_2\oplus 1-\rho(0)$ 
to an orthogonal basis of the traceless matrices $\cA_0=V+\bR z$. We 
have
\begin{align}
\label{eq:coor}
\langle E(\alpha,s,t),\sigma_1\oplus 0\rangle & \; = \;
\tfrac 1{\eta}\big[2\sinh(b)x/b\big]\\\nonumber
\langle E(\alpha,s,t),\sigma_2\oplus 1\rangle & \; = \;
\tfrac 1{\eta}
\big[2\sinh(b)y/b+e^{-s\sin(\alpha)+t\cos(\alpha)}\big]\\\nonumber
\langle E(\alpha,s,t),0_2\oplus 1-\rho(0)\rangle & \; = \;
\tfrac 1{\eta}
\big[-\cosh(b)-\sinh(b)y/b+e^{-s\sin(\alpha)+t\cos(\alpha)}\big]\,.
\end{align}
\par
We discuss closures of the Staffelberg family and its entropy distance.
\begin{Thm}
\label{thm:closure}
The Staffelberg family $\cE$ has $(+1)$-closure and rI-closure equal 
to ${\rm cl}^{(+1)}(\cE)={\rm cl}^{\rm rI}(\cE)=\cE\cup B\cup\{c\}$.
The norm closure is
$\overline{\cE}={\rm cl}^{\rm rI}(\cE)\cup[\rho(0),c]$. The entropy 
distance of $\rho\in[\rho(0),0_2\oplus 1]$ from $\cE$ is 
${\rm d}_\cE(\rho)=S(\rho,c)$. The restricted projection 
$\pi_V|_{{\rm cl}^{\rm rI}(\cE)}$ is a bijection onto the mean value 
set $\bM(V)$.
\end{Thm}
{\em Proof:\/} 
By Proposition~\ref{pro:geodesic_closure} the $(+1)$-closure of $\cE$
is a union of exponential families 
$\cE^q=\left\{\exp_1^q(q\theta q)\mid\theta\in V\right\}$ for 
maximal projections $q$. In place of the maximal projections of 
$v\neq 0$ in $V$ we consider equivalently the maximal projections of 
the vectors
\begin{equation}\textstyle
u(\alpha)\;:=\;\sin(\alpha)\sigma_1\oplus 0 +
\cos(\alpha)(\sigma_2\oplus 1)\,,
\qquad\alpha\in\bR\,.
\end{equation}
There are two cases depending on the spectral projections in the 
orthogonal sum
\[\textstyle
u(\alpha)\;=\;\rho(\alpha)-\rho(\alpha+\pi)
+0_2\oplus\cos(\alpha)\,.
\]
The maximal eigenvalue of $u(\alpha)$ is constant one. If
$\alpha\neq 0\;{\rm mod}\;2\pi$, then the maximal projection of $u(\alpha)$
is $\rho(\alpha)$ and has rank one. We get
\[
\cE^{\rho(\alpha)}
\;=\;
\left\{\exp_1^{\rho(\alpha)}({\rho(\alpha)}\theta{\rho(\alpha)})
\mid\theta\in V\right\}
\;=\;
\{\rho(\alpha)\}
\]
proving $B\subset{\rm cl}^{(+1)}(\cE)$. If $\alpha=0\;{\rm mod}\;2\pi$, 
then the maximal projection of $u(0)$ is $p:=\rho(0)+0_2\oplus 1=2c$. 
Since $p(\sigma_1\oplus 0)p=0$ and $p(\sigma_2\oplus 1)p=p$ the canonical
parameter space of $\cE^p$ consists of multiples of the identity $p$
in $p\cA p$, $p\Theta p=pVp=\bR p$. So is $\cE^p=\{c\}$, we conclude 
${\rm cl}^{(+1)}(\cE)=\cE\cup B\cup\{c\}$.
\par
Lemma~\ref{lem:upper_bound} provides an upper bound on the norm closure
$\overline{\cE}$ in terms of faces of $\st(\cA)$ exposed by vectors 
in $V$, and Lemma~\ref{lem:exposed_of_state} describes these faces 
in terms of maximal projections
\[\textstyle
F(\st(\cA),u(\alpha))
\;=\;
\{\rho\in\st(\cA)\mid s(\rho)\preceq p^+(u(\alpha))\}\,.
\]
For $\alpha\neq 0\;{\rm mod}\;2\pi$ the maximal projection $\rho(\alpha)$ 
of $u(\alpha)$ has rank one and the exposed face is 
$F(\st(\cA),u(\alpha))=\{\rho(\alpha)\}$. The projection
$p^+(u(0))=p$ above gives the segment 
$[\rho(0),0_2\oplus 1]=F(\st(\cA),u(0))$. We obtain
\[
\overline{\cE}\;\subset\;\cE\,\cup\,B\,\cup\,[\rho(0),0_2\oplus 1]\,.
\]
The inclusions ${\rm cl}^{(+1)}(\cE)\subset\overline{\cE}$ and
$B\subset{\rm cl}^{(+1)}(\cE)$ show $B\subset\overline{\cE}$. We prove 
that exactly the part $[\rho(0),c]$ of the segment 
$[\rho(0),0_2\oplus 1]$ belongs to $\overline{\cE}$.
\par
We prove that at most the half segment $[\rho(0),c]$ belongs to 
$\overline{\cE}$ by showing that $\cE$ is included in
the closed half space $\langle a,v_3\rangle\leq 0$. This is sufficient
because $v_3$ is parallel to $[\rho(0),0_2\oplus 1]$ and
\[\textstyle
\langle \rho(0),v_3\rangle\;=\;-1\,,\qquad
\langle c,v_3\rangle\;=\;0\,\qquad
\textrm{and}\qquad
\langle 0_2\oplus 1,v_3\rangle\;=\;1\,.
\]
We look at the polar parametrization of $\cE$, defined with
(\ref{eq:def_polar_ste}) as
\[
\bR\times\bR_0^+\to\cE,\quad
(\alpha,t)\mapsto E(\alpha,0,t)\,.
\]
The normalization factor $\eta$ is strictly positive, so 
$\langle E(\alpha,0,t),v_3\rangle\leq 0$ is by (\ref{eq:coor})  
equivalent to 
\[\textstyle
z(\alpha,t)
\; := \; \eta \langle E(\alpha,0,t),v_3\rangle
\; = \; -\cos(\alpha)\sinh(t)-\cosh(t)+e^{\cos(\alpha)t}
\; \leq \;0\,.
\]
For $t=0$ we have $z(\alpha,0)=0$ while for $t\geq 0$ and 
arbitrary $\alpha\in\bR$ we have
\[\textstyle
\pm z(\alpha,t)+\frac{\partial}{\partial t}z(\alpha,t)
\; = \;
(\cos(\alpha)\pm 1)\left[e^{\cos(\alpha)t}-e^{\pm t}\right]
\; \leq \; 
0\,.
\]
This implies $\frac{\partial}{\partial t}z(\alpha,t)\leq 0$ and by 
integration $z(\alpha,t)\leq 0$.
\par
We show $[\rho(0),c]\subset\overline{\cE}$. The state $\rho(0)$ lies in 
the closure of $B$ so we still have to approximate for $\lambda\in(0,1]$ 
the state
$\tau(\lambda):=(1-\frac{\lambda}{2})\rho(0)\oplus\frac{\lambda}{2}$ 
from within $\cE$. For $t>0$ we choose 
$\alpha(t):=\sqrt{\frac{2}{t}\ln(\frac{2-\lambda}{\lambda})}$. 
Then $\lim_{t\to\infty}\alpha(t)=0$ and
$\lim_{t\to\infty}e^{(\cos(\alpha(t))-1)t}=\frac{\lambda}{2-\lambda}$
hold. Expanding by $e^{-t}$ we have
\[\textstyle
\lim_{t\to\infty}E(\alpha(t),0,t)\;=\;
\frac{\frac{1}{2}(\id_2+\sigma_2)\oplus\frac{\lambda}{2-\lambda}}{%
1+\frac{\lambda}{2-\lambda}}
\;=\;\tau(\lambda)\,.
\]
\par
We calculate the rI-closure. This is bounded by 
Corollary~\ref{cor:inclusions} between $(+1)$- and norm closures
\[
{\rm cl}^{(+1)}(\cE)
\;=\;\cE\cup B\cup\{c\}
\;\subset\;{\rm cl}^{\rm rI}(\cE)
\;\subset\;\cE\cup B\cup[\rho(0),c]
\;=\;\overline{\cE}\,.
\]
It remains to discuss states 
$\rho\in[\rho(0),0_2\oplus 1]=F(\st(\cA),u(0))$. 
Proposition~\ref{pro:inclusion} and $\cE^p=\{c\}$ show 
\[\textstyle
{\rm d}_\cE(\rho)={\rm d}_{\cE^p}(\rho)=S(\rho,c)\,.
\]
So $\rho\in{\rm cl}^{\rm rI}(\cE)$ holds for 
$\rho\in[\rho(0),0_2\oplus 1]$ if and only if $\rho=c$.
This shows ${\rm cl}^{\rm rI}(\cE)={\rm cl}^{(+1)}(\cE)$.
\par
We show that $\pi_V|_{{\rm cl}^{\rm rI}(\cE)}$ is a bijection onto
$\bM(V)$. The boundary of the mean value set $\bM(V)$ is by 
(\ref{eq:mean_set}) and by Lemma~\ref{Lem:nonex} equal to the ellipse 
\[\textstyle
\partial\bM(V)=\pi_V(B\cup\{\rho(0)\})
\]
so $\pi_V$ restricted to the circle $B\cup\{\rho(0)\}$ is a bijection. 
Since $c$ lies on the segment $[\rho(0),0_2\oplus 1]$ which is 
perpendicular to $V$, it substitutes $\rho(0)$ in that bijection.
Another bijection is the mean value chart 
$\pi_V|_{\cE}\;:\cE\to\ri({\bM}(V))$, see (\ref{eq:mean_chart}). 
The two latter bijections assembled prove the claim.
\hspace*{\fill}$\Box$\\
\begin{Cor}
\label{cor:disc}
The entropy distance ${\rm d}_\cE:\st(\cA)\to[0,\log(3)]$ from the 
Staffelberg family is discontinuous at $\rho(0)$.
\end{Cor}
{\em Proof:\/} 
By the previous theorem we have ${\rm d}_\cE(\rho(0))=S(\rho(0),c)=\ln(2)$ 
while ${\rm d}_\cE\equiv 0$ on the punctured base circle $B$ of the cone 
$\mathcal S(\mathcal A)$. But $\rho(0)\in\overline{B}$.
\hspace*{\fill}$\Box$\\
\begin{Cor}
\label{cor:disc_ext}
The mean value parametrization 
$\pi_\cE:{\rm ri}(\bM(V))\to\cE$ of the Staffelberg family has no 
continuous extension to the mean value set $\bM(V)$; it has no
continuous extension to $\pi_V(\rho(0))$.
\end{Cor}
{\em Proof:\/} 
Since the segment $[\rho(0),0_2\oplus 1]$ belongs to the norm closure of 
$\cE$ and since this segment is perpendicular to $V$, the mean value 
parametrization $\pi_\cE:{\rm ri}(\bM(V))\to\cE$ does not extend 
continuously to $\pi_V(\rho(0))$.
\hspace*{\fill}$\Box$\\
\par
We address the maximum-entropy principle. 
\begin{Thm}
\label{thm:max_entropy_family}
The rI-closure of the Staffelberg family is a set of maximum-entropy 
density matrices,
${\rm cl}^{\rm rI}(\cE)=\{{\rm argmax}_{\rho\in F(v)}S(\rho)\mid
v\in\bM(V)\}$.
This holds for fibers $F(v):=(v+V^\perp)\cap\st(\cA)$ as well as for
\[F(v)\;:=\;(v+V^\perp)\;\cap\;\st({\rm Mat}(3,\mathbb C)).\]
\end{Thm}
{\em Proof:\/} 
Since the Staffelberg family $\cE$ is included in the state space
$\mathcal S(\mathcal A)$, the Pinsker-Csisz\'ar inequality, recalled 
in Corollary~\ref{cor:inclusions}, shows that $\cE$ has the same 
rI-closure in both algebras $\mathcal A$ and ${\rm Mat}(3,\mathbb C)$.
The mean value chart (\ref{eq:mean_chart}) shows that the mean value
set $\mathbb M(V)$ is the same for both algebras. So
the bijection $\pi_V|_{{\rm cl}^{\rm rI}(\cE)}$ from the rI-closure 
onto the mean value set, proved in Theorem~\ref{thm:closure}, also 
applies to both algebras.
\par
We discuss the inverse $\bM(V)\to{\rm cl}^{\rm rI}(\cE)$.
Its restriction to the interior of the mean value set
$\ri\big({\bM}(V)\big)\to\cE$ is the mean value parametrization 
of $\cE$ and this 
is known to have the maximum-entropy property (\ref{eq:mean_par_inference}).
\par
Let us now consider the boundary of the mean value set $\bM(V)$, which is
by (\ref{eq:mean_set}) and by Lemma~\ref{Lem:nonex} equal to the ellipse 
\[\textstyle
\partial\bM(V)=\pi_V\big(B\cup\{\rho(0)\}\big)\,.
\]
The fibers $F(\widetilde{v})$ for points 
$\widetilde{v}\in\partial\bM(V)$ are faces of the state space
$\st(\cA)$, see Section~5 in \cite{Weis_touch}. Indeed they are the
set of state space faces $F(\st(\cA),v)$ which are exposed
by a non-zero $v\in V$. Using Lemma~\ref{lem:exposed_of_state}
and consulting the list of maximal projections of vectors $v\in V$ in 
Theorem~\ref{thm:closure} these faces are the points on the punctured
circle $B$ and the segment 
$[\rho(0),0_2\oplus 1]$. Maximizers of the 
von Neumann entropy on these fibers are the points on $B$ and the 
centroid $c$ in the segment. This set completes $\cE$ to its rI-closure 
by Theorem~\ref{thm:closure}.
\par
In the larger C*-algebra ${\rm Mat}(3,{\bC})$ the projection
$\rho(0)+0_2\oplus 1$ corresponds to the face
$\{\rho\in\st({\rm Mat}(3,\bC))\mid s(\rho)\preceq \rho(0)+0_2\oplus 1\}$
which is isomorphic to the Bloch ball. So the maximizer of the
von Neumann entropy in the fiber $(v+V^\perp)\cap\st({\rm Mat}(3,{\bC}))$
is $c$ as before.
\hspace*{\fill}$\Box$\\
\par
We finish with two short conclusions about a discontinuous inference.
\begin{Rem}
\label{rem:max_entropy_family}
If a maximum-entropy inference (\ref{eq:maxent_inference}) is carried 
out by observables spanning the canonical tangent space $V$ of the 
Staffelberg family, then the variance of the inferred state 
$\widehat{\rho}(n)$ may be large: Assuming that the quantum system is 
given by an invertible density matrix $\rho$, measured values 
$\big(m_1,\ldots,m_k\big)$ are mapped to the inferred state 
$\widehat{\rho}(n)$ by the mean value parametrization $\pi_\cE$
defined in (\ref{eq:mean_par_inference}). The mean value parametrization
$\pi_\cE$ does not extend continuously to $\pi_V(\rho(0))$ by 
Corollary~\ref{cor:disc_ext} so the mean value theorem shows that
$\pi_\cE$ has arbitrary large partial  derivatives near $\pi_V(\rho(0))$.
It follows that the constant in the variance estimate 
${\cal O}(1/n)$ of $\widehat{\rho}(n)$ can be arbitrarily large.
\par
Second, the non-generic choice of $\rho$ such that
$\pi_V(\rho)=\pi_V(\rho(0))$ makes it likely that the inferred 
states $\widehat{\rho}(n)$ diverge or converge to a state 
which is not a maximum-entropy state. This follows from 
Theorem~\ref{thm:closure} and Theorem~\ref{thm:max_entropy_family} 
because the whole segment $[\rho(0),c]$ belongs to the closure of 
$\cE$ while only $c$ is a state of maximum entropy under the given
constraints.
\end{Rem}
%
%%%%%%%%%%%%%%%%%%%%%%%%%%%%%%%%%%%%%%%%%%%%%%%%%%%%%%%%%%%%%%%%%%%%%%%%
%%%%%%%%%%%%%%%%%%%%%%%%%%%%%%%%%%%%%%%%%%%%%%%%%%%%%%%%%%%%%%%%%%%%%%%%
%%%%%%%%%%%%%%%%%%%%%%%%%%%%%%%%%%%%%%%%%%%%%%%%%%%%%%%%%%%%%%%%%%%%%%%%
%
%
\subsection{$(+1)$-asymptotics and $(-1)$-closure of the Staffelberg 
family}
\label{sec:-1-staffelberg}
\par
We show that the $(-1)$-closure of the Staffelberg family $\cE$ 
equals its rI-closure. This follows from an asymptotic analysis of
its $(+1)$-geodesics. See 
(\ref{eg:1_closure}) and (\ref{eg:-1_closure})
for definitions of these closures.
\par
We use the parametrization $E(\alpha,s,t)$ of $\cE$ defined in 
(\ref{eq:def_polar_ste}) and a coordinate system spanned by
$(\sigma_2\oplus 1)$ and $(\sigma_1\oplus 0)$. Coefficients of points
on $\cE$ are the first two numbers in (\ref{eq:coor}), they 
describe projection onto $V$: 
\begin{align*}
g & \;:=\;
\langle E(\alpha,s,t),\sigma_2\oplus 1\rangle
\;=\;
\tfrac 1{\eta}
\big[(e^b-e^{-b})y/b+e^{-s\sin(\alpha)+t\cos(\alpha)}\big]\\
h & \;:=\;
\langle E(\alpha,s,t),\sigma_1\oplus 0\rangle
\;=\;
\tfrac 1{\eta}\big[(e^b-e^{-b})x/b\big]\,.
\end{align*}
We consider the asymptotic slope in the 
$(\sigma_2\oplus 1)$-$(\sigma_1\oplus 0)$-coordinate system
\begin{equation}
\label{eq:asy_slope}
\kappa(\alpha,s)
\;:=\;
\lim_{t\to\infty}\frac{{\rm d}h}{{\rm d}g}
\;=\;
\lim_{t\to\infty}\frac{
\tfrac{{\rm d}h}{{\rm d}t}}{
\tfrac{{\rm d}g}{{\rm d}t}}
\;=\;
\lim_{t\to\infty}
\frac{\eta \tfrac{{\rm d}(h\eta)}{{\rm d}t} 
- (h\eta) \tfrac{{\rm d}\eta}{{\rm d}t}}{%
\eta \tfrac{{\rm d}(g\eta)}{{\rm d}t} 
- (g\eta) \tfrac{{\rm d}\eta}{{\rm d}t}}\,.
\end{equation}
The coordinates 
$\{(\langle\rho,\sigma_2\oplus 1\rangle,
\langle\rho,\sigma_1\oplus 0\rangle)\mid\rho\in\st(\cA)\}$ of the mean 
value set fill the unit disk. 
Projections of $(+1)$-geodesics hit the unit circle for $s=0$, they are 
tangential to the unit circle for every $s\neq 0$:
\begin{Lem}
\label{lem:(+1)-asymptotics}
For all $\alpha\in\bR$ and all $s\in\bR$ we have
$(g,h)\stackrel{t\to\infty}{\longrightarrow}(\cos(\alpha),\sin(\alpha))$.
The asymptotic slope of $(+1)$-geodesics through the tracial state
($s=0$) is
\[
\kappa(\alpha,0)
\;=\;
\left\{\begin{array}{rcl} 0 & \text{if} & \alpha=0\,,\\
-\cot(\tfrac{\alpha}{2}) & \text{if} & \alpha\neq 0\,.
\end{array}\right.
\]
The asymptotic slope of $(+1)$-geodesics missing the tracial state
($s\neq 0$) is
\[
\kappa(\alpha,s)
\;=\;
-\cot(\alpha)\,.
\]
\end{Lem}
{\em Proof:\/}
The $(+1)$-geodesic limit $t\to\infty$ follows from (\ref{asy:exp1}) 
and from the discussion of maximal projections in 
Theorem~\ref{thm:closure}. Then $\lim_{t\to\infty}(g,h)$ follows.
\par
We first compute the asymptotical slope for $(+1)$-geodesics through
the tracial state $s=0$. We have
\begin{align*}
& (\eta \tfrac{{\rm d}(h\eta)}{{\rm d}t} 
- (h\eta) \tfrac{{\rm d}\eta}{{\rm d}t})
e^{-t(1+\cos(\alpha))}\\
& =
\sin(\alpha)(1 + e^{-2 t} + 
 4 e^{-t - t \cos(\alpha)} - 
\cos(\alpha) + e^{-2 t}\cos(\alpha))
\end{align*}
and
\begin{align*}
& (\eta \tfrac{{\rm d}(g\eta)}{{\rm d}t} 
- (g\eta) \tfrac{{\rm d}\eta}{{\rm d}t})
e^{-t(1+\cos(\alpha))}\\
& =
-(1 - \cos(\alpha))^2 + e^{-2 t} + \cos(\alpha)( 2 e^{-2 t} + 
 4 e^{-t - t\cos(\alpha)} + e^{-2 t} \cos(\alpha))\,.
\end{align*}
From this and (\ref{eq:asy_slope}) we get the desired result, 
studying $\alpha=0$ and $\alpha=\pi$ apart.
\par
The asymptotical slope for $(+1)$-geodesics missing the tracial 
state $(s\neq 0)$ follows from a third order Taylor expansion at 
$t=\infty$. If $\alpha\neq 0$ modulo $2\pi$ then
\begin{align*}
(\eta \tfrac{{\rm d}(h\eta)}{{\rm d}t} 
- (h\eta) \tfrac{{\rm d}\eta}{{\rm d}t})
& \;=\;
-\frac{s}{t^2}\cos(\alpha)+O(\tfrac{1}{t^3})\\
(\eta \tfrac{{\rm d}(g\eta)}{{\rm d}t} 
- (g\eta) \tfrac{{\rm d}\eta}{{\rm d}t})
& \;=\;
\frac{s}{t^2}\sin(\alpha)+O(\tfrac{1}{t^3})\,.
\end{align*}
For $\alpha=0$ we have
\begin{align*}
(\eta \tfrac{{\rm d}(h\eta)}{{\rm d}t}
- (h\eta) \tfrac{{\rm d}\eta}{{\rm d}t})
& \;=\;
-\frac{2s}{t^2}+O(\tfrac{1}{t^3})\\
(\eta \tfrac{{\rm d}(g\eta)}{{\rm d}t} 
- (g\eta) \tfrac{{\rm d}\eta}{{\rm d}t})
& \;=\;
O(\tfrac{1}{t^3})
\end{align*}
completing the claim.
\hspace*{\fill}$\Box$\\
\begin{figure}
\begin{center}
\begin{picture}(12,5)
\ifthenelse{\equal{\eps}{ja}}{%
\put(0,0){\includegraphics[height=5cm, bb=0 0 300 297, clip=]%
{eGeo_hit.eps}}
\put(7,0){\includegraphics[height=5cm, bb=0 0 300 297, clip=]%
{eGeo_parallel.eps}}}{%
\put(0,0){\includegraphics[height=5cm, bb=0 0 355 355, clip=]%
{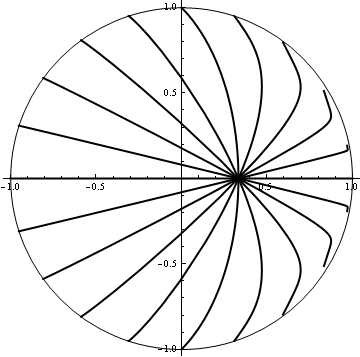}}
\put(7,0){\includegraphics[height=5cm, bb=0 0 355 355, clip=]%
{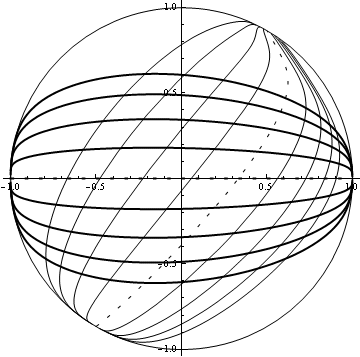}}}
%
%\put(0,0){\line(1,0){5}}
%\put(0,5){\line(1,0){5}}
%\put(0,0){\line(0,1){5}}
%\put(5,0){\line(0,1){5}}
%
%\put(7,0){\line(1,0){5}}
%\put(7,5){\line(1,0){5}}
%\put(7,0){\line(0,1){5}}
%\put(12,0){\line(0,1){5}}
\end{picture}
\caption{\label{fig:(+1)-geodesics}
Projected $(+1)$-geodesics in the Staffelberg family. Left: 
geodesics through the tracial state; right: two families of 
parallel geodesics, those through the tracial state are dashed.}
\end{center}
\end{figure}
\par
Some projected $(+1)$-geodesics of the Staffelberg family are drawn 
in Figure~\ref{fig:(+1)-geodesics}. As a fact not used in the sequel, 
Lemma~\ref{lem:(+1)-asymptotics} shows that the two asymptotic tangents 
$t\to\pm\infty$ of a projected $(+1)$-geodesic through the tracial 
state ($s=0$) intersect orthogonally at $(1,0)$ for $\alpha\neq 0,\pi$.
While the right angle is not invariant under affine reparametrizations, 
these tangents intersect in $V$ at the projection of the cliff 
$c=\tfrac{1}{2}(\rho(0)+0_2\oplus 1)$
of the Staffelberg family.
\begin{Lem}
\label{lem:uniform_convergence}
For all $s\in[-1,1]$ and all $t\geq 1$ we have uniformly in $s$
\[
\|E(0,s,t)-c\|_2={\cal O}(t^{-1})\,.
\]
\end{Lem}
{\em Proof:\/}
By Taylor expansion $b=t+ \frac{s^2}{2t}+{\cal O}(t^{-2})$, we have
uniformly for $s\in[-1,1]$
\[{\textstyle
E(0,s,t)
\;=\;
\bsm \cosh(b)& (s-i t)\frac{\sinh(b)}{b}&0\\
 (s+i t)\frac{\sinh(b)}{b}& \cosh(b)&0\\
0&0&e^t\esm \!\! \Big/\big(2\cosh(b)+e^t\big)
\;=\;
c+{\cal O}(t^{-1})}\,.\]
This  proves the statement, since 
$\|a\|_2=\sqrt{\sum_{k,\ell} |a_{k,\ell}|^2}$.
\hspace*{\fill}$\Box$\\
\begin{Thm}
\label{thm:staffel+1}
For the Staffelberg family $\cE$ the $(-1)$-closure equals the
$(+1)$- and the rI-closure, 
${\rm cl}^{(-1)}(\cE)={\rm cl}^{(+1)}(\cE)={\rm cl}^{\rm rI}(\cE)$.
\end{Thm}
{\em Proof:\/}
The equality ${\rm cl}^{(+1)}(\cE)={\rm cl}^{\rm rI}(\cE)$ was shown
in Theorem~\ref{thm:closure}. Since $(-1)$-geodesics are included 
in $\cE$ we clearly have ${\rm cl}^{(-1)}(\cE)\subset\overline{\cE}$.
On the other hand, in every fiber $(v+V^\perp)\cap\st(\cA)$ with
$v\in\bM(V)$ there is at least one point of the $(-1)$-closure
(choose a segment $]u,v[\subset{\rm ri}(\bM(V))$ and lift it to $\cE$
through the mean value parametrization). By Theorem~\ref{thm:closure}
there is a bijection
\[
\pi_V|_{\overline{\cE}\setminus S}:
\overline{\cE}\setminus S\to\bM(V)\setminus\{m\}
\]
for the segment $S:=[\rho(0),0_2\oplus 1]$ and its projection
$m:=\pi_V(c)$. The three arguments combined show
$\overline{\cE}\setminus S={\rm cl}^{(-1)}(\cE)\setminus S$.
\par
It remains to discuss states $\rho\in S$, whether they belong to 
${\rm cl}^{(-1)}(\cE)$. The point $c$ clearly does since the 
unparametrized $(-1)$-geodesic $]\rho(\pi),c[$ belongs to $\cE$. 
We finish by showing $\{c\}=S\cap{\rm cl}^{(-1)}(\cE)$.
\par
The $(-1)$-geodesic from $\rho(\pi)$ to $c$ is also a $(+1)$-geodesic, 
parametrized for $s=0$ by
\[
g_s(t)
\;:=\;
E(0,s,t)\,.
\]
Using (\ref{asy:exp1}) we see that for all real $s$ the geodesic $g_s$ 
has the limit $c$ when $t\to+\infty$, its projection $\pi_V(g_s)$ has
the limit $m=\pi_V(c)$. For $s\neq 0$ the asymptotic tangent of 
$\pi_V(g_s)$ is tangential to the elliptical boundary $\partial\bM(V)$ 
of the mean value set by Lemma~\ref{lem:(+1)-asymptotics}. This implies 
that the projections $\pi_V(g_{-1})$ and $\pi_V(g_{+1})$ concatenate 
to a closed smooth curve in $\bM(V)$ which is tangential to
$\partial\bM(V)$ at $m$. Using the mean value chart 
(\ref{eq:mean_chart}) of $\cE$, it is clear that this curve bounds the 
set
\[
U
\;:=\;
\{\pi_V(g_s(t))\mid -1\leq s\leq 1, t\in\bR\}
\;\subset\;
\bM(V)\,.
\]
\par
Let $h$ be any $(-1)$-geodesic in $\cE$ with limit $\rho$ in the 
segment $S$. If we choose any sequence $\rho_n\subset h$ such that 
$\rho=\lim_{n\to\infty}\rho_n$, then $\theta_n:=\log_0(\rho_n)$ 
diverges in the norm (otherwise the contradiction $\rho\in\cE$ 
follows). As the boundary of $U$ is tangential to the ellipse 
$\partial\bM(V)$ at $m$, there is $\epsilon>0$ such that
\[
\pi_V(h)\cap\{v\in V\mid \|v-m\|_2<\epsilon\}
\;\subset\;
U\,.
\]
So the points $\pi_V(\rho_n)$ lie in $U$ for large $n$. Since the 
convergence of the $(+1)$-geodesics $g_s$ to $c$ is uniform (for 
$-1\leq s\leq 1$) by Lemma~\ref{lem:uniform_convergence}, the states 
$\rho_n$ converge to $c$.
\hspace*{\fill}$\Box$\\
%
%
%
%%%%%%%%%%%%%%%%%%%%%%%%%%%%%%%%%%%%%%%%%%%%%%%%%%%%%%%%%%%%%%%%%%%%%%%%
%%%%%%%%%%%%%%%%%%%%%%%%%%%%%%%%%%%%%%%%%%%%%%%%%%%%%%%%%%%%%%%%%%%%%%%%
%%%%%%%%%%%%%%%%%%%%%%%%%%%%%%%%%%%%%%%%%%%%%%%%%%%%%%%%%%%%%%%%%%%%%%%%
%
\subsection{The Swallow family}
\label{sec:swallow}
\par
\begin{figure}
\begin{center}
\begin{picture}(5.5,6)
\ifthenelse{\equal{\eps}{ja}}{%
\put(0,0){\includegraphics[height=6cm, bb=60 40 495 448, clip=]%
{Fig51.eps}}}{%
\put(0,0){\includegraphics[height=6cm, bb=60 40 495 448, clip=]%
{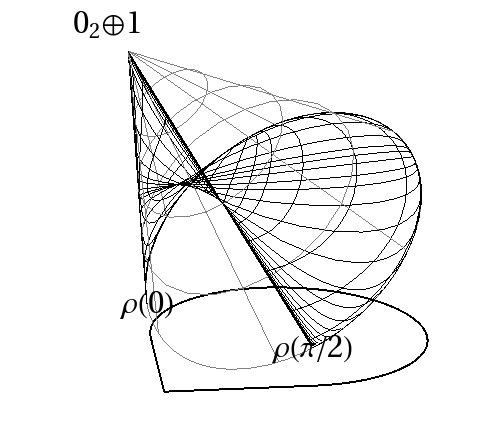}}}
%\put(0,0){\line(1,0){5.5}}
%\put(0,6){\line(1,0){5.5}}
%\put(0,0){\line(0,1){6}}
%\put(5.5,0){\line(0,1){6}}
\end{picture}
\caption{\label{fig:swallow}
The Swallow family $\cE$ sketched by $(+1)$-geodesics. The cone about 
$\cE$ is the state space $\st(\cA)$. Its generating lines 
$[\rho(0),0_2\oplus 1]$ and $[\rho(\frac{\pi}{2}),0_2\oplus 1]$ belong 
to the rI-closure of $\cE$ but the pure states $\rho(0)$ and 
$\rho(\frac{\pi}{2})$ do not belong to the $(+1)$-closure of $\cE$.
They project to the non-exposed points of the mean value set $\bM(V)$
whose boundary is drawn below.}
\end{center}
\end{figure}

We now consider 2D families $\cE = \exp_1(V)$ in the metamorphosis of 
Figure~\ref{fig:reflections} that have non-exposed faces in the mean 
value set $\bM(V)$. By Lemma~\ref{Lem:nonex} this happens for
angles $\varphi(V)\in(0,\pi/3)$. 
We prove that the $(+1)$-closure ${\rm cl}^{(+1)}(\cE)$ is too small 
to serve as a set of entropy maximizers under linear constraints. The
problem is that the two non-exposed points of the mean value set are 
not covered by ${\rm cl}^{(+1)}(\cE)$ in the projection onto $V$.
Calculations become easy for
$\varphi={\rm arccos}(\sqrt{2/5})\approx0.28\pi$ and we then call $\cE$
the \emph{Swallow family} because it looks like the beak of a bird:
\begin{Def}
\label{def:swallow}
The \emph{Swallow family}, depicted in Figure~\ref{fig:swallow}, 
is the Gibbsian family
\[
\cE\;:=\;\exp_1\left({\rm span}_\bR\{\sigma_1\oplus 1,
\sigma_2\oplus 1\}\right)
\]
in the *-subalgebra $\cA\subset{\rm Mat}(3,{\mathbb C})$ 
defined in Example~\ref{exa:algebra}. 
\end{Def}
\par
The canonical tangent space $V=\Theta$ of $\cE$ is spanned by 
the vectors of equal length
$v_1:=\sigma_1\oplus 1-\frac{1}{3}\id$ and
$v_2:=\sigma_2\oplus 1-\frac{1}{3}\id$. The vector 
$z=-\frac{1}{2}\id_2\oplus 1$ is perpendicular to $v_2-v_1$, so
indeed
\[\textstyle
\varphi=\winkel(V,z)=\winkel(v_1+v_2,z)={\rm arccos}(\sqrt{2/5})\,.
\]
The pure states $\rho(0)=\frac{1}{2}(\id_2+\sigma_2)\oplus 0$ and 
$\rho(\frac{\pi}{2})=\frac{1}{2}(\id_2+\sigma_1)\oplus 0$ on the base 
circle of the conic state space $\st(\cA)$ are crucial for the
Swallow family.
\begin{Thm}
\label{thm:swallow}
The $(+1)$-closure of the Swallow family $\cE$ is the union of $\cE$, 
of the segments $]\rho(0),0_2\oplus 1[$ and
$]\rho(\frac{\pi}{2}),0_2\oplus 1[$ (rank-two states) and of the pure 
states $0_2\oplus 1$ and $\{\rho(\alpha)\mid\frac{\pi}{2}<\alpha<2\pi\}$.
The $(-1)$- rI- and norm closures are
\[\textstyle
{\rm cl}^{(-1)}(\cE)={\rm cl}^{\rm rI}(\cE)=\overline{\cE}
\;=\;
{\rm cl}^{(+1)}(\cE)\,\cup\,\{\rho(0),\rho(\frac{\pi}{2})\}\,.
\]
\end{Thm}
{\em Proof:\/}
First we calculate the
$(+1)$-closure ${\rm cl}^{(+1)}(\cE)$
using Proposition~\ref{pro:geodesic_closure}. For $\alpha\in\bR$ we have 
the orthogonal sum
\[\textstyle
u(\alpha)\;:=\;\sin(\alpha)(\sigma_1\oplus 1) +
\cos(\alpha)(\sigma_2\oplus 1)
\;=\;\rho(\alpha)-\rho(\alpha+\pi)+0_2\oplus\sqrt{2}
\cos(\alpha-\frac{\pi}{4})\,.
\]
The maximal projections for $\alpha=0$ and $\frac{\pi}{2}$ 
are
\[\textstyle
p\;:=\;p^+(u(0))=\rho(0)+0_2\oplus 1\,\quad\text{and}\quad
q\;:=\;p^+(u({\scriptstyle \frac{\pi}{2}}))
= \rho({\scriptstyle \frac{\pi}{2}})+0_2\oplus 1\,.
\]
For $0<\alpha<\frac{\pi}{2}$ we have
$p^+(u(\alpha))=0_2\oplus 1$ and for $\frac{\pi}{2}<\alpha<2\pi$ we have
$p^+(u(\alpha))=\rho(\alpha)$.
\par
Calculating the corresponding exponential families we observe
$p\cA p\cong\bC^2$ and since $p(\sigma_1\oplus 1)p=0_2\oplus 1$, 
the exponential family $\cE^p=\exp_1^p(p\Theta p)$ has the canonical 
parameter space
\[\textstyle
\bR(0_2\oplus 1-\rho(0))
\;\cong\;
\bR(1,-1)\subset\bC^2\,.
\]
The analogue arguments apply to $q$, so the exponential family  
\[\textstyle
\cE^p\;=\;\,]\rho(0),0_2\oplus 1[
\quad\textrm{resp.}\quad
\cE^q\;=\;\,]\rho(\frac{\pi}{2}),0_2\oplus 1[
\]
consists of the invertible states in the compressed algebra $p\cA p$ 
resp.\ $q\cA q$. All other maximal projections 
$r$ of elements of $v\neq 0$ of $V$ have rank one and produce the
exponential family
$\cE^r=\left\{\exp_1^r(r\theta r)\mid\theta\in V\right\}=\{r\}$. 
This completes the calculation of the 
$(+1)$-closure of $\cE$.
\par
In the second step we prove that the points $\rho(0)$ and 
$\rho(\frac{\pi}{2})$ missing in the $(+1)$-closure belong to the 
rI-closure of $\cE$. Lemma~\ref{lem:exposed_of_state} describes 
the exposed face $F(\st(\cA),u(0))=[\rho(0),0_2\oplus 1]=\cS(p\cA p)$, 
containing the pure state $\rho(0)$. Then  
Proposition~\ref{pro:inclusion} shows
\[\textstyle
{\rm d}_\cE(\rho(0))
\;=\;
{\rm d}_{\cE^p}(\rho(0))
\;=\;
{\rm d}_{{\rm cl}^{(+1)}(\cE^p)}(\rho(0))
\,.
\]
Since
\[\textstyle
{\rm cl}^{(+1)}(\cE^p)
\;=\;
{\rm cl}^{(+1)}(\,]\rho(0),0_2\oplus 1[\,)
\;=\;
[\rho(0),0_2\oplus 1]
\]
we get ${\rm d}_\cE(\rho(0))={\rm d}_{[\rho(0),0_2\oplus 1]}(\rho(0))=0$
and this implies $\rho(0)\in{\rm cl}^{\rm rI}(\cE)$. The
analogue arguments show $\rho(\frac{\pi}{2})\in{\rm cl}^{\rm rI}(\cE)$.
\par
By the same method as in Theorem~\ref{thm:closure} an upper bound on 
the norm closure $\overline{\cE}$ can be stated in terms of maximal 
projections in $V$. These projections are listed above, the
corresponding faces
are the pure state $0_2\oplus 1$, the arc of pure states $\rho(\alpha)$
for $\frac{\pi}{2}<\alpha<2\pi$ and the two segments 
$[\rho(0),0_2\oplus 1]$ and $[\rho(\frac{\pi}{2}),0_2\oplus 1]$
(the state spaces of the algebras $p\cA p\cong q\cA q\cong\bC^2$). 
Thus $\overline{\cE}\subset
{\rm cl}^{(+1)}(\cE)\cup\{\rho(0),\rho(\frac{\pi}{2})\}$ follows from
the above description of the $(+1)$-closure. Since $\rho(0)$ and
$\rho(\frac{\pi}{2})$ belong to the rI-closure and since
${\rm cl}^{(+1)}(\cE)\subset{\rm cl}^{\rm rI}(\cE)\subset\overline{\cE}$
holds by Corollary~\ref{cor:inclusions} we have shown
${\rm cl}^{\rm rI}(\cE)=\overline{\cE}=
{\rm cl}^{(+1)}(\cE)\cup\{\rho(0),\rho(\frac{\pi}{2})\}$.
\hspace*{\fill}$\Box$\\
\begin{Thm}
\label{thm:swallow2}
The projection $\pi_V|_{{\rm cl}^{\rm rI}(\cE)}$ is a bijection onto 
the mean value set $\bM(V)$, the non-exposed points of $\bM(V)$ are 
$\pi_V(\rho(0))$ and $\pi_V(\rho(\frac{\pi}{2}))$. The rI-closure of 
the Swallow family is a set of maximum-entropy density matrices,
\[\textstyle
{\rm cl}^{\rm rI}(\cE)=\{{\rm argmax}_{\rho\in F(v)}S(\rho)\mid
v\in\bM(V)\}
\]
for fibers $F(v):=(v+V^\perp)\cap\st(\cA)$.
\end{Thm}
{\em Proof:\/}
The relative interiors of faces of the mean value set $\bM(V)$ are
a partition of $\bM(V)$ \cite{Rockafellar}. Each face $F$ of $\bM(V)$ 
is the projection to $V$ of the inverse projection 
$(F+V^\perp)\cap\st(\cA)$, which is a face of $\st(\cA)$. The 
relative interior of the inverse projection of $F$ projects onto
the relative interior of $F$; we show that these projections are
bijections for the algebra $\cA$, for the Swallow family $\cE$ and
for all faces $F$ in the boundary of the mean value set $\bM(V)$.
\par
The two non-exposed points 
$\pi_V(\rho(0))$ and $\pi_V(\rho(\frac{\pi}{2}))$ at the ellipse with 
corner $\bM(V)$ are computed in case 3 of Example 1.2 in 
\cite{Weis_supp} studying tangents. 
The present setting fits into 
Example 1.2 in \cite{Weis_supp} by choosing there 
$g:=\tfrac 1{\sqrt{2}}(1,-1,0)$ and $h:=\tfrac 1{\sqrt{2}}(1,1,0)$.
The inverse projections $(\rho(0)+V^\perp)\cap\st(\cA)$ and 
$(\rho(\tfrac{\pi}{2})+V^\perp)\cap\st(\cA)$ are faces of the 
state space $\st(\cA)$ and it is proved in case 3 of Section 3.3 in
\cite{Weis_supp} that these faces are the extremal 
points $\rho(0)$ and $\rho(\tfrac{\pi}{2})$ and that they are not 
larger.
\par
Every exposed face $F=F(\bM(V),v)$ for non-zero $v\in V$ is actually 
the projection of the exposed face $F(\st(\cA),v)$, see Section 3.1 
in \cite{Weis_supp}. These faces are computed in the last paragraph
of Theorem~\ref{thm:swallow}. A missing bijectivity of their 
projections onto $V$ is only possible for the two segments, but it 
does not occur because the two segments cover the two boundary segments 
of $\bM(V)$.
\par
The maximum-entropy problem is solved for points in 
${\rm ri}(\bM(V))$ in (\ref{eq:mean_par_inference}). Since the
projection of $(\partial\bM(V)+V^\perp)\cap\st(\cA)$ onto $V$ is 
a bijection onto $\partial\bM(V)$, the maximum-entropy problem is 
trivial for boundary points of $\bM(V)$.
\hspace*{\fill}$\Box$\\
\begin{Rem}
\label{rem:swallow}
\begin{enumerate}[a)]
\item
The Swallow family is suitable to demonstrate that the extreme points
of a mean value set $\bM(V)$ are in general not covered by the
projections $\pi_V(\tfrac p{\tr(p)})$ for maximal projections
$p=p^+(v)$, $v\in V$, as is claimed in Theorem~1~(e) in \cite{Wichmann}.
\par
Let $\mathcal B$ denote one of the algebras $\cA$ or 
${\rm Mat}(3,\mathbb C)$ where $\cA\subset{\rm Mat}(3,{\mathbb C})$
is the *-subalgebra  defined in Example~\ref{exa:algebra}.
Since $\cA$ and ${\rm Mat}(3,\mathbb C)$ 
have the same identities $\id=\id_3$ we can argue with eigenvalues 
to calculate the maximal projections of vectors in $V$. Moreover, the 
mean value set $\mathbb M_\cB(V)$ is well-defined, see Section 3.4 
in \cite{Weis_supp}. For faces $F$ of the mean value set the lifted 
faces $(F+V^\perp)\cap\cS(\mathcal B)$ are of the form 
$\{\rho\in\cS(\mathcal B)\mid s(\rho)\preceq p\}$ for projections 
$p\in\mathcal B$, see Section 2.3 in \cite{Weis_supp}. The necessary 
projections $p$ are computed recursively from $V$, see Theorem 3.7 
or Remark 3.10 in \cite{Weis_supp}. This gives the same set of 
projections for both algebras $\mathcal A$ and ${\rm Mat}(3,\mathbb C)$.
\par
Now, the pure state $\rho(0)$ (and $\rho(\tfrac{\pi}{2})$) is not on the 
list of maximal projections of vectors in $V$ provided in the first 
paragraph of Theorem~\ref{thm:swallow}. On the other hand, 
as discussed in the second paragraph of Theorem~\ref{thm:swallow2}, 
the state $\rho(0)$ is the unique
state in $\cS(\cA)$ that projects to the non-exposed point 
$\pi_V(\rho(0))$ of the mean value set.
\item
There is no $(+1)$-geodesic in the Swallow family $\cE$ that meets 
$\rho(0)$ asymptotically. Calculation of ${\rm cl}^{\rm rI}(\cE)$ in
Theorem~\ref{thm:swallow} is done by two limits of $(+1)$-geodesics. 
One of the limits is implicit in the equation 
${\rm d}_\cE(\rho(0))={\rm d}_{\cE^p}(\rho(0))$. Only a second 
$(+1)$-geodesic in $\cE^p$ meets $\rho(0)$ asymptotically.
\end{enumerate}
\end{Rem}
%
%
%
%
%
%%%%%%%%%%%%%%%%%%%%%%%%%%%%%%%%%%%%%%%%%%%%%%%%%%%%%%%%%%%%%%%%%%%%%%%%
%%%%%%%%%%%%%%%%%%%%%%%%%%%%%%%%%%%%%%%%%%%%%%%%%%%%%%%%%%%%%%%%%%%%%%%%
%%%%%%%%%%%%%%%%%%%%%%%%%%%%%%%%%%%%%%%%%%%%%%%%%%%%%%%%%%%%%%%%%%%%%%%%
%
%
%
\section{Maximizers of the entropy distance}
\label{sec:maxi}
\par
We now study local maximizers of the entropy distance ${\rm d}_\cE$ 
from an exponential family $\cE$, a question which was motivated in 
Section~\ref{max_inf} in the context of infomax principles.
We have to restrict to Gibbsian families since the mean value chart 
(\ref{eq:mean_chart}) is only available for these exponential families 
in the present article.
\par
We show that a local maximizer $\rho$ of ${\rm d}_\cE$ carries a 
clear imprint from its projection $\pi_{\cE}(\rho)$ to $\cE$. This 
generalizes the commutative case, where $\rho$ is the conditional 
probability distribution of $\pi_\cE(\rho)$ conditioned on its own 
support ${\rm supp}(\rho)$
\begin{equation}
\label{eq:cl_imprint}
\rho\;=\;\pi_{\cE}(\rho)(\;\cdot\;|{\rm supp}(\rho))\,.
\end{equation}
\begin{Rem}
In the commutative case the assertion (\ref{eq:cl_imprint}) was 
proved for a local maximizer 
$\rho\in{\rm dom}\cE=\st(\cA)\cap(\cE+V^{\perp})$ in \cite{Ay}.
The articles \cite{Knauf,Matus07,Rauh,Matus_Rauh}
contain further characterizations of local and global
maximizers that can be interesting also in the non-commutative case.
\end{Rem}
\par
The derivative of the logarithm is derived for $\cA={\rm Mat}(N,\bC)$
in \cite{Lieb}. It may be generalized to any *-subalgebra $\cA$ 
of ${\rm Mat}(N,\bC)$ using an algebra embedding 
$\phi:\cA\to{\rm Mat}(n,\bC)$ such that $\phi(\id)$ is invertible.
If $p\in\cA$ is a projection then for invertible $\rho\in\st(p\cA p)$ 
and self-adjoint $u\in p\cA p$ we have
\begin{equation}
\label{eq:deriv_log}\textstyle
{\rm D}|_\rho\ln^p(u)
\;=\;\int_0^{\infty}(\rho+s p)^{-1}u(\rho+s p)^{-1}{\rm d}s\,.
\end{equation}
Here we denote functions in $p\cA p$ by a superscript like in the 
paragraph before Lemma~\ref{lem:e-limits}.
\begin{Thm}
\label{thm:local_max}
Suppose $\cA$ is a *-subalgebra  of ${\rm Mat}(N,{\mathbb C})$ and 
$\cE$ a Gibbsian family in $\cA$ with canonical tangent space $V$.
Let $\rho\in{\rm dom}\,\cE$, let $p$ 
denote the support projection of $\rho$ and put 
$\theta:=\ln_0\circ\pi_\cE(\rho)\in V$. If $u$ is a traceless 
self-adjoint matrix in $p\cA p$, then 
${\rm D}|_{\rho}{\rm d}_\cE(u)=\langle u,\ln^p(\rho)-\theta\rangle$. 
If $\rho$ is a local maximizer of ${\rm d}_{\cE}$, then 
$\rho=\exp_1^p(p\,\theta p)$ and 
${\rm d}_{\cE}(\rho)=F(\theta)-F^p(p\,\theta p)$.
\end{Thm}
{\em Proof:\/} 
As discussed in the paragraph following (\ref{eq:mean_chart}), the 
mean value parametrization $\pi_{\cE}$ defined for $a\in\cE+V^{\perp}$ 
by intersection $a\mapsto(a+V^{\perp})\cap\cE$ is real analytic. This 
gives a real analytic mapping
\[\textstyle
L\;:\quad\cE+V^{\perp}\,\longrightarrow\,V,
\quad a\,\longmapsto\,\ln_0\circ\,\pi_{\cE}(a)\,.
\]
We can use $\pi_{\cE}(a)=\exp_1\circ L(a)$ and rewrite the entropy 
distance (\ref{eq:entropy_distance}) of a state $\rho\in\cE+V^{\perp}$ 
from $\cE$ in the form
\begin{eqnarray}
\label{eq:d_expansion}\textstyle
\lefteqn{{\rm d}_{\cE}(\rho)\;=\;S(\rho,\pi_{\cE}(\rho))
\; = \; S(\rho,\exp_1\circ L(\rho))}\\\nonumber
& = & -S(\rho)-\tr(\rho\ln\circ \exp_1\circ L(\rho))
\; = \; -S(\rho)-\tr(\rho L(\rho))+F\circ L(\rho)
\hspace{0.5cm}
\end{eqnarray}
with the free energy $F$ and von Neumann entropy $S$. As $\rho$ is 
invertible in the algebra $p\cA p$, we can differentiate at $\rho$ 
the logarithm $\ln^p$ in the direction of any self-adjoint matrix 
$u\in p\cA p$. By (\ref{eq:deriv_log}) and cyclic reordering under 
the trace we get
\[\textstyle
{\rm D}|_{\rho}S(u)
\;=\;-\langle u,\ln^p(\rho)\rangle-\tr(u)\,.
\]
Using the derivative of the free energy (\ref{eq:free_derivative}), 
which is for $a,b\in\cA$ given by
${\rm D}|_a F(b)\;=\;\langle b,\exp_1(a)\rangle$, the chain rule leads 
to
\begin{eqnarray*}
\lefteqn{{\rm D}|_{\rho}(F\circ L)(u)
\; = \; {\rm D}|_{L(\rho)}F\circ{\rm D}|_{\rho}L(u)}\\
& = & \langle{\rm D}|_{\rho}L(u),\exp_1\circ L(\rho)\rangle
\;=\;\langle{\rm D}|_{\rho}L(u),\pi_{\cE}(\rho)\rangle\,.
\end{eqnarray*}
Since the image of $L$ is $V$ we have ${\rm D}|_{\rho}L(u)\in V$ and thus 
by definition of the projection $\pi_{\cE}$ follows
$\langle {\rm D}|_{\rho}L(u),\pi_{\cE}(\rho)-\rho\rangle=0$. 
Differentiation of (\ref{eq:d_expansion}) in the direction of a 
traceless self-adjoint matrix $u\in p\cA p$ gives
\begin{eqnarray*}
\lefteqn{{\rm D}|_{\rho}{\rm d}_{\cE}(u)
\;=\;\langle u,\ln^p(\rho)\rangle + \tr(u)
-\langle u,L(\rho)\rangle
-\langle\rho,{\rm D}|_{\rho}L(u)\rangle}\\
& & +\langle{\rm D}|_{\rho}L(u),\pi_{\cE}(\rho)\rangle
\; = \; \langle u,\ln^p(\rho)-L(\rho)\rangle\,.
\hspace{2cm}
\end{eqnarray*}
This completes the asserted directional derivative.
\par
If $\rho$ is a local maximizer of ${\rm d}_{\cE}$, then
$\ln^p(\rho)=p\,L(\rho)p+\lambda p$ for some real $\lambda$ because $p$ 
spans the orthogonal complement of the space of traceless self-adjoint 
matrices in $p\cA p$. If follows that $\rho$ must be proportional to 
$p\exp(p\,L(\rho)p)$ as claimed. If we write
$\theta:=L(\rho)=\ln_0\circ\pi_{\cE}(\rho)$, then we have
$\rho=\exp_1^p(p\,\theta p)$ and $\pi_{\cE}(\rho)=\exp_1(\theta)$. We get 
\begin{eqnarray*}
\lefteqn{{\rm d}_{\cE}(\rho)
\; = \; S(\rho,\pi_\cE(\rho))
\; = \; \tr[\rho(\ln^p(\rho)-\ln\circ\pi_{\cE}(\rho))]}\\
 & = &  
\tr\left[\rho(p\,\theta p-p\ln\circ\tr\circ\exp^p(p\,\theta p)
-\theta+\id\ln\circ\tr\circ\exp(\theta))\right]\\
& = & \ln(\tr(e^{\theta}))-\ln(\tr(p\,e^{p\,\theta p}))\,.
\end{eqnarray*}
\hspace*{\fill}$\Box$\\
\par
{\textbf{\emph{Acknowledgment:}} SW thanks the organizers of the DFG 
research group ``Geometry and Complexity in Information Theory'' 
(2004--2008) for the scholarship and the great workshops.
We thank Nihat Ay for discussions about information measures
and the referee for several helpful comments.
\bibliographystyle{nnapalike}

\end{document}